\documentclass[vecphys]{svmult}

\usepackage{makeidx}         
\usepackage{graphicx}        
\usepackage{multicol}        
\usepackage{cite}            
\usepackage[bottom]{footmisc}

\usepackage{color,soul}      
\usepackage{amsmath}
\usepackage{textgreek}
\usepackage{setspace}

\makeindex             

\begin{document}

\title{A Microscopic and Spectroscopic View of Quantum Tunneling of Magnetization}
\titlerunning{A Microscopic and Spectroscopic View of QTM}
\author{Junjie Liu \and Enrique del Barco \and Stephen Hill}
\institute{Department of Physics, University of Florida, Gainesville, Florida 32611, USA \texttt{jliu@magnet.fsu.edu}
\and Department of Physics, University of Central Florida, Orlando, Florida 32816, USA \texttt{delbarco@physics.ucf.edu}
\and Department of Physics and National High Magnetic Field Laboratory, Florida State University, Tallahassee, Florida 32310, USA \texttt{shill@magnet.fsu.edu}}

\maketitle

\begin{abstract}This chapter takes a microscopic view of quantum tunneling of magnetization (QTM) in single-molecule magnets (SMMs), focusing on the interplay between exchange and anisotropy. Careful consideration is given to the relationship between molecular symmetry and the symmetry of the spin Hamiltonian that dictates QTM selection rules. Higher order interactions that can modify the usual selection rules are shown to be very sensitive to the exchange strength. In the strong coupling limit, the spin Hamiltonian possess rigorous $D_{2h}$ symmetry (or $C_{\infty}$ in high-symmetry cases). In the case of weaker exchange, additional symmetries may emerge through mixing of excited spin states into the ground state. Group theoretic arguments are introduced to support these ideas, as are extensive results of magnetization hysteresis and electron paramagnetic resonance measurements.

\end{abstract}

\noindent
\section{Spin Hamiltonian}
\label{sec:SH}
The concept of an effective spin-Hamiltonian involving only spin variables has been employed in the study of paramagnetic species for well over half a century. This formalism is particularly suited to the study of transition metal complexes in which the ground state is very often an orbital singlet that is well isolated from excited orbital states due to the strong influence of the ligand field~\cite{Rudowicz2001ASR}. The ground state multiplicity is determined entirely by the spin state of the ion in this situation, even though the ground wave functions are not exact eigenstates of $\bf {\hat S}^2$ because of residual spin-orbit (SO) coupling. It is this coupling that gives rise to the familiar anisotropic zero-field splitting (zfs) and Zeeman terms in the resultant spin Hamiltonian, parameterized by the zfs ${\buildrel{\lower3pt\hbox{$\scriptscriptstyle\leftrightarrow$}}\over D}$ and Lande ${\buildrel{\lower3pt\hbox{$\scriptscriptstyle\leftrightarrow$}}\over g}$-tensors.
\subsection{Giant-Spin Approximation Hamiltonian}
\label{sec:GSA} 
The magnetic moment of a typical polynuclear transition metal cluster is determined by the exchange interactions between the spins associated with the constituent ions. As detailed in this chapter, there are a number of ways to extend the spin Hamiltonian formalism to this multi-ion situation. By far the simplest is the so-called Giant Spin Approximation (GSA), in which one assigns a total (giant) spin quantum number, $S$, to the lowest-lying ($m_s$) magnetic levels~\cite{Gatteschi2006Oxford}; for a ferromagnetic molecule, $S$ is obtained from the algebraic sum of the spin values associated with each of the ions. If the exchange coupling within the molecule is large in comparison to the single-ion zfs interactions, then this ground spin multiplet will be well separated from excited spin states. One may then employ a GSA Hamiltonian to describe the magnetic properties of the molecule, provided that the temperature is sufficiently low that excited spin states are not thermally populated.

A series expansion in terms of the spin component operators $\hat{S}_{x}$, $\hat{S}_{y}$, and $\hat{S}_{z}$, employing so-called Extended Stevens operators, results in the following effective zfs Hamiltonian\cite{Abragam&Bleaney1986Dover,Rudowicz2004JPCM,Stoll2006JMR}:
\begin{equation}
\label{eqn:GSA}
\hat{H}_{\text{zfs}}=\sum\limits_p^{2S} \sum\limits_{q=0}^p B_{p}^{q}\hat{O}_{p}^{q},
\end{equation}
where $\hat{O}_p^q(\hat{S}_{x},\hat{S}_{y},\hat{S}_{z})$ represent the operators, and $B_p^q$ the associated phenomenological (or effective) zfs parameters. The subscript, $p$, denotes the order of the operator, which must be even due to the {time reversal invariance} of the SO interaction; the order is also limited by the total spin, $S$, of the molecule such that $p \leq 2S$. The superscript, $q$ ($\leq p$), denotes the rotational symmetry of the operator about the \textit{z}-axis. Eqn.\,\ref{eqn:GSA} has been employed with great success in the study of single-molecule magnets (SMMs), particularly in terms of describing low-temperature quantum tunneling of magnetization (QTM) behavior and electron paramagnetic resonance (EPR) data~\cite{Gatteschi2006Oxford}. In fact, Eqn.\,\ref{eqn:GSA} has even been applied quite successfully in cases where the ground spin multiplet is not so well isolated from excited spin states~\cite{Hill2009PRB,Quddusi2011,Liu2011,Heroux2011}. Fourth and higher order operators are often found to be important in these cases. This chapter examines the microscopic origin of these terms, which are often negligible for the constituent ions. However, the usually dominant 2\textsuperscript{nd}-order zfs interaction is first considered.

\subsubsection{{Second Order Anisotropy}}
Although fourth and higher order Stevens operators ($p\geq 4$) are allowed for a molecule with $S\geq 2$, it is not always necessary to include all of them (up to $p=2S$). For SMMs comprised of transition metal ions, the low-energy physics is usually dominated by 2\textsuperscript{nd}-order SO anisotropy. Therefore, the simplest zero-field GSA Hamiltonian used in characterizing SMMs is often written as:

\begin{equation}
\label{eqn:GSA2nd}
\hat{H}_{\text{zfs}}=D\hat{S}_z^2+E(\hat{S}_x^2-\hat{S}_y^2).
\end{equation}
Eqn.\,\ref{eqn:GSA2nd} includes only 2\textsuperscript{nd}-order terms, where $D$ ($=3B_2^0$) parameterizes the {uniaxial anisotropy} and $E$ ($=B_2^2$) the rhombicity. For an approximately uniaxial system, $D\hat{S}_z^2$ is the dominant anisotropy, with \textit{z} chosen as the quantization axis. In biaxial cases, the ratio between \textit{E} and \textit{D} is usually restricted such that $|E/D|<1/3$; one can always perform a rotation of the coordinate system such that this criterion is satisfied.

\begin{figure}[t]
\centering
\includegraphics*[width=.7\textwidth]{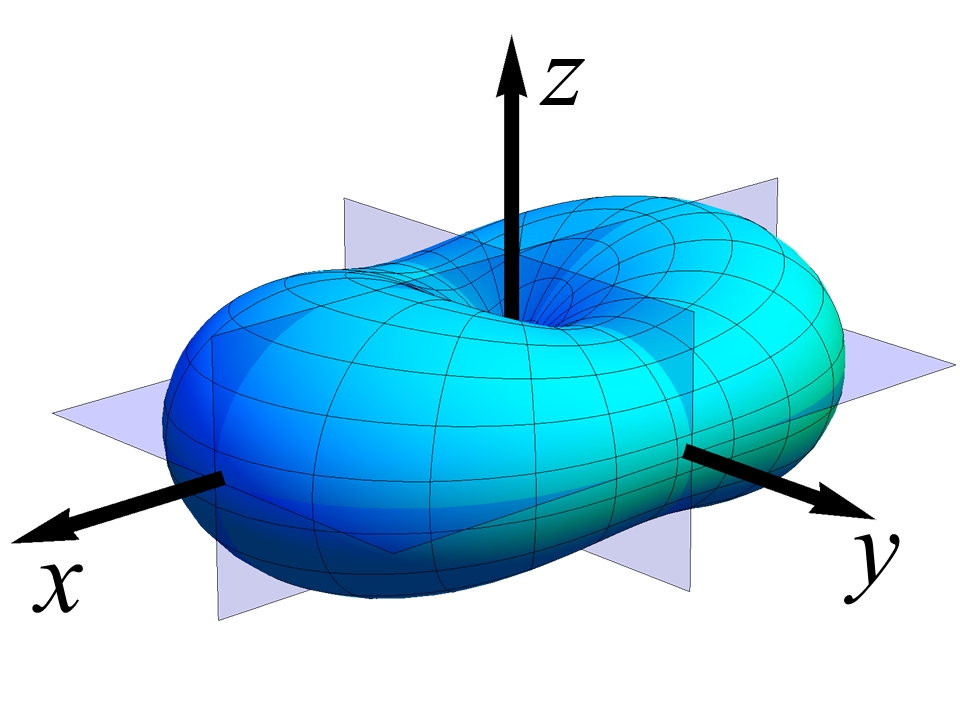}
\caption[]{Potential energy surface corresponding to the 2\textsuperscript{nd}-order anisotropy tensor. The surface is generated employing Eqn.\,\ref{eqn:GSA2nd} with $|E/D| = 1/5$ and $D < 0$. The radial distance to the surface represents the energy of a spin as a function of its orientation.}
\label{fig:GSA2nd}
\end{figure}

One of the main goals of this section is to understand the influence of molecular symmetry on the QTM properties of SMMs. Hence, it is important to examine the symmetry of Eqn.\,\ref{eqn:GSA2nd} since, strictly speaking, the symmetry of the Hamiltonian should be compatible with the symmetry of the molecule under investigation. In addition, the nature of SO coupling ensures that the spin Hamiltonian be invariant under time-reversal (\textit{p} is even), \textit{i.e.}, the spin Hamiltonian naturally possesses $C_i$ symmetry. As such, the physics is invariant to inversion of either the total spin moment or the applied field. A classical representation of Eqn.\,\ref{eqn:GSA2nd} is shown in Fig.\,\ref{fig:GSA2nd}, with $|E|=|D|/5$ and $D<0$. This graphical representation is obtained by substituting the spin operators in Eqn.\,\ref{eqn:GSA2nd} by their classical equivalents, as follows:
\begin{equation}
\label{eqn:classicalspin}
\begin{aligned}
\hat{S}_x&\rightarrow S\sin\theta\cos\phi;\\
\hat{S}_y&\rightarrow S\sin\theta\sin\phi;\\
\hat{S}_z&\rightarrow S\cos\theta;
\end{aligned}
\end{equation}
where $\theta$ and $\phi$ are the inclination and azimuthal angles in spherical coordinates, respectively. In this representation, the spin is treated as a macroscopic magnetic moment for which all the three components ($S_x$, $S_y$ and $S_z$) can be determined simultaneously. The surface shown in Fig.\,\ref{fig:GSA2nd} represents the energy of the spin as a function of its orientation, where the radial distance to the surface corresponds to its energy. As can be seen, Eqn.\,\ref{eqn:GSA2nd} contains the following symmetry elements: (a) three orthogonal $C_2$ axes, corresponding to \textit{x}, \textit{y} and \textit{z}, and (b) three orthogonal mirror planes, corresponding to the \textit{xy}, \textit{yz}, and \textit{zx}-planes. These symmetry elements, together with the {$C_i$ symmetry}, give rise to a {$D_{2h}$ symmetry} for Eqn.\,\ref{eqn:GSA2nd}, which is obviously a much higher symmetry than most real molecules. Consequently, even though Eqn.\,\ref{eqn:GSA2nd} can often account very well for low-temperature thermodynamic measurements performed on SMMs (\textit{e.g.} ac susceptibility and magnetization), it may nevertheless fail to explain symmetry-sensitive quantum mechanical (spectroscopic) observables such as QTM steps, Berry-phase interference (BPI) patterns and EPR/neutron spectra.

It should be noted from the preceding discussion that the $B_2^1\hat{O}_2^1$ $(\hat{O}_2^1 \equiv \frac{1}{2}[\hat{S}_x \hat{S}_z + \hat{S}_z \hat{S}_x])$ term was neglected in Eqn.\,\ref{eqn:GSA2nd}, even though it is perfectly allowed within the GSA. With the exception of $C_i$, the $\hat{O}_2^1$ operator satisfies none of the symmetry operations described in the preceding paragraph. However, such a term is unnecessary, as can be seen when writing the 2\textsuperscript{nd}-order GSA Hamiltonian in the more compact form:
\begin{equation}
\label{eqn:GSA2ndTensor}
\hat{H}_{\text{zfs}}=\vec{\hat{S}}\cdot \buildrel{\lower3pt\hbox{$\scriptscriptstyle\leftrightarrow$}}\over{D} \cdot \vec{\hat{S}},
\end{equation}
where $\buildrel{\lower3pt\hbox{$\scriptscriptstyle\leftrightarrow$}}\over{D}$ is a $3\times 3$ matrix corresponding to the full 2$^\textrm{nd}$-order anisotropy tensor. In Eqn.\,\ref{eqn:GSA2ndTensor}, \textit{D} and \textit{E} are related to the diagonal elements of $\buildrel{\lower3pt\hbox{$\scriptscriptstyle\leftrightarrow$}} \over D$ (see below) while $B_2^1$ appears as off-diagonal elements. The only restriction on $\buildrel{\lower3pt\hbox{$\scriptscriptstyle\leftrightarrow$}} \over D$ is that it must be Hermitian in order to guarantee the Hamiltonian be Hermitian; indeed, $D_{xz} = D_{zx} = \frac{1}{2} B_2^1$. Consequently, $\buildrel{\lower3pt\hbox{$\scriptscriptstyle\leftrightarrow$}} \over D$ can always be diagonalized by rotating the original Cartesian coordinate frame. Upon doing so, all of the off-diagonal elements of the rotated matrix vanish, \textit{i.e.}, $B_2^1 = 0$ in the new Cartesian coordinate frame. Finally, one may adjust the absolute values of the resultant eigenvalues without altering the symmetry of the Hamiltonian simply by subtracting $\frac{1}{2}(D_{xx}+D_{yy})$$\buildrel{\lower3pt\hbox{$\scriptscriptstyle\leftrightarrow$}}\over I$ from $\buildrel{\lower3pt\hbox{$\scriptscriptstyle\leftrightarrow$}} \over D$ ($\buildrel{\lower3pt\hbox{$\scriptscriptstyle\leftrightarrow$}} \over I$ is the identity matrix). The zfs Hamiltonian can then be rewritten as Eqn.\,\ref{eqn:GSA2nd} with
\begin{equation}
\label{eqn:GSA2ndTransform}
D=D_{zz}-\frac{1}{2}(D_{xx}+D_{yy}) \quad \textrm{and} \quad E=\frac{1}{2}(D_{xx}-D_{yy}),
\end{equation}
where $D_{ii}$ ($i=x,y,z$) refer to components of the diagonalized (rotated) $\buildrel{\lower3pt\hbox{$\scriptscriptstyle\leftrightarrow$}}\over{D}$ tensor. In other words, Eqn.\,\ref{eqn:GSA2ndTensor} is equivalent to Eqn.\,\ref{eqn:GSA2nd}, requiring just two parameters, $D$ and $E$, to completely describe the effective 2\textsuperscript{nd}-order anisotropy within the GSA. Inclusion of $\hat{O}_2^1$ results simply in a rotation of the surface depicted in Fig.\,\ref{fig:GSA2nd}. Consequently, the 2\textsuperscript{nd}-order GSA Hamiltonian necessarily possesses at least $D_{2h}$ symmetry.

The preceding discussion demonstrates something very important: even though 2\textsuperscript{nd}-order terms typically represent the dominant interactions within the GSA description of a SMM, the resultant $\buildrel{\lower3pt\hbox{$\scriptscriptstyle\leftrightarrow$}} \over D$ tensor possesses an artificially high ($D_{2h}$) symmetry which may not be compatible with the structural symmetry of a particular molecule under investigation. The consequences of this property of the GSA in terms of the resultant QTM will be discussed in detail in following sections.

\subsubsection{{Higher-Order Anisotropies}}
\label{sec:HighOrder}

For a SMM with $S\geq2$, Stevens operators of order four (and higher) are allowed in the GSA Hamiltonian$\--$note $p$ can take on any even value from 2 to $2S$. The values of the 4\textsuperscript{th}-order parameters are often deceptively small, especially for SMMs with large spin values. For example, $|B^0_4/D|\sim5\times10^{-5}$ for the Mn$_{12}$ SMMs, yet the $B^0_4\hat{O}_4^0$ GSA term contributes $\sim20\%$ to the {energy barrier}. This is due not only to the higher order of $\hat{S}_z$ in $\hat{O}_4^0$, but also because of the way in which the $\hat{O}_4^0$ operator is defined$\--$a multiplier of 35 is associated with $S_z^4$. In general, the contribution of higher-order terms to the energies of spin states may be expected to be smaller than those of the 2\textsuperscript{nd}-order terms. However, this rule of thumb breaks down in the {weak exchange limit} (or for particularly high-symmetry molecules~\cite{Beedle2013IC}); indeed, it is in this limit that one may call into question the validity and/or usefulness of the GSA. Axial ($q=0$) 4\textsuperscript{th}-order terms lead to a non-parabolic energy barrier, which gives rise to non-even spacings between EPR and QTM resonance fields \cite{Hill1998PRL,Barra2000CEJ,Takahashi2004PRB,Kirman2005JAP}. More importantly, the higher-order transverse ($q\not= 0$) terms introduce additional symmetries into the GSA Hamiltonian, enabling a more precise description of the quantum properties of SMMs.

Fig.\,\ref{fig:GSAHighOrder} displays the {classical energy surfaces} corresponding to the {4\textsuperscript{th}-order Stevens operators}; the $\hat{O}_4^0$ surface is not shown since it commutes with $\hat{S}_z$ and possesses $C_{\infty}$ (cylindrical) rotational symmetry. All of the surfaces, and hence the operators, exhibit rotational symmetries which are compatible with the superscript \textit{q}. However, one may note a systematic difference between the \textit{q}-odd and \textit{q}-even operators: the \textit{q}-even operators have the \textit{xy}-plane as an additional mirror plane; this is not the case for the \textit{q}-odd operators. In particular, the lobes of the $\hat{O}_4^2$ and $\hat{O}_4^4$ operators lie within the \textit{xy}-plane, whereas those of $\hat{O}_4^1$ and $\hat{O}_4^3$ alternately lie above and below this plane. The symmetries of the 4\textsuperscript{th}-order operators can be understood in terms of combinations of rotational symmetries and the intrinsic $C_i$ symmetry of the Hamiltonian. For the \textit{q}-even operators, the direct product of the rotational and inversion group leads to: $C_2 \times C_i = C_{2h}$ symmetry for $\hat{O}_4^2$; and $C_4 \times C_i = C_{4h}$ for $\hat{O}_4^4$. Thus, the \textit{xy}-plane is introduced as a new symmetry element. In contrast, for the \textit{q}-odd operators, $C_1 \times C_i = C_i$ for $\hat{O}_4^1$ and $C_3 \times C_i = S_6$ for $\hat{O}_4^3$. The resultant symmetry groups corresponding to these operators include an {improper rotation} ($C_i$ can be treated as the improper rotation $S_2$). The absence of the \textit{xy}-mirror plane for the \textit{q}-odd operators suggests that the molecular hard plane may not coincide with the \textit{xy}-plane, which leads to several intriguing phenomena described later in this chapter.

The inclusion of $p\geq4$ Stevens operators in the GSA has a significant influence on the interpretation of {QTM measurements}. When limited to 2\textsuperscript{nd}-order anisotropy, the zero-field Hamiltonian can only mix spin projection states that differ in $m_s$ by an even number, \textit{i.e.}, $\text{\textDelta} m_s = |m_{s1} - m_{s2}|=2n$ ($n=$integer). This means that only $k$-even ($k=m_{s1}+m_{s2}$) QTM resonances should be observable for parallel applied fields ($H\parallel z$). Moreover, in molecules for which rhombicity is symmetry forbidden ($E=0$), a purely 2\textsuperscript{nd}-order Hamiltonian would be cylindrically symmetric. Consequently, for $H\parallel z$, $m_s$ remains an exact quantum number and QTM should be completely forbidden. However, plenty of such high-symmetry molecules exist and are known to exhibit clear QTM behavior, including Mn$_{12}$ and several other SMMs discussed in this chapter. In these situations, it is necessary to include higher order anisotropies in the GSA. The corresponding operators introduce new spin-mixing rules which can lead, for example, to $k$-odd QTM resonances.

The advantage of the GSA lies in the fact that one can usually restrict the total number of zfs parameters involved in data analysis to just a few by considering the overall symmetry of the molecule under study. Furthermore, the GSA Hilbert space includes only the $2S+1$ states that belong to the ground spin multiplet, such that the Hamiltonian matrix has dimension $(2S+1)\times(2S+1)$. This makes data analysis for large clusters computationally possible. However, the GSA completely ignores the internal degrees of freedom within a molecule, thus completely failing to capture the underlying physics in cases where the total spin fluctuates \cite{Carretta2004PRL,Wilson2006PRB,Ramsey2008NaturePhys}. Moreover, when a molecule possesses very little symmetry (\textit{e.g.} $C_i$), the number of GSA zfs parameters cannot be restricted on the basis of symmetry and, in principle, all possible terms (up to $p=2S$) should be taken into account. In these cases, it may be advantageous to employ a multi-spin Hamiltonian, particularly in situations where microscopic insights are desired.

\subsection{{Multi-Spin Hamiltonian}}
\label{sec:MS}

In the multi-spin (MS) model, a molecule is treated as a cluster of magnetic ions (spins) which are coupled to each other via pairwise exchange interactions. The corresponding zero-field Hamiltonian is:

\begin{equation}
\label{eqn:MS}
\hat{H}_{\text{zfs}}=\sum\limits_i \vec{\hat{s}}_i \cdot {\mathord{\buildrel{\lower3pt\hbox{$\scriptscriptstyle\leftrightarrow$}} \over R}{} _i^T} \cdot \buildrel{\lower3pt\hbox{$\scriptscriptstyle\leftrightarrow$}}\over{d}_i \cdot \buildrel{\lower3pt\hbox{$\scriptscriptstyle\leftrightarrow$}} \over R_i \cdot \vec{\hat{s}}_i + \sum\limits_{i<j} \vec{\hat{s}}_i \cdot \buildrel{\lower3pt\hbox{$\scriptscriptstyle\leftrightarrow$}} \over J_{i,j} \cdot \vec{\hat{s}}_j,
\end{equation}

\noindent{where} $\vec{\hat{s}}_i$ represents the spin operator of the $i^{\text{th}}$ ion, and $\buildrel{\lower3pt\hbox{$\scriptscriptstyle\leftrightarrow$}} \over d_i$ is the 2\textsuperscript{nd}-order zfs tensor associated with this same ion; lowercase symbols are used here to differentiate parameters/variables employed in both models, \textit{i.e.}, lowercase $\equiv$ MS and uppercase $\equiv$ GSA. For the sake of simplicity, $\buildrel{\lower3pt\hbox{$\scriptscriptstyle\leftrightarrow$}}\over{d}_i$ is written in the diagonal form:

\begin{equation}
\label{eqn:2ndMatrix}
\buildrel{\lower3pt\hbox{$\scriptscriptstyle\leftrightarrow$}} \over d~=~\left[ \begin{array}{ccc}
e_i & 0 & 0 \\
0 & -e_i & 0 \\
0 & 0 & d_i \\
\end{array} \right],
\end{equation}

\noindent{where} the local coordinate frame of $\buildrel{\lower3pt\hbox{$\scriptscriptstyle\leftrightarrow$}} \over d_i$ is chosen to match the local principal anisotropy axes of the $i^{\textrm{th}}$ ion. $\buildrel{\lower3pt\hbox{$\scriptscriptstyle\leftrightarrow$}} \over R_i$ is the Euler matrix, specified by the Euler angles $\theta_i$, $\phi_i$ and $\psi_i$, which transforms the local coordinate frame of the $i^{\textrm{th}}$ ion into the molecular coordinate frame. The matrix $\buildrel{\lower3pt\hbox{$\scriptscriptstyle\leftrightarrow$}} \over J_{i,j}$ specifies the {exchange interaction} between the $i^{\text{th}}$ and $j^{\text{th}}$ ions. It should be emphasized that all of the parameters in Eqn.\,\ref{eqn:MS} should be constrained by the structure of the molecule under study, \textit{i.e.}, the overall symmetry of the Hamiltonian must be compatible with the {molecular symmetry}.

The MS model captures physics associated with internal molecular degrees of freedom that are not easily understood within the GSA framework. First and foremost, the MS model is capable of describing phenomena in which the total spin of a molecule fluctuates, \textit{i.e.}, it gives the energies of {excited spin states} in addition to the ground state, and includes the mixing between these states~\cite{Hill2010Dalton}. Secondly, the parameters in the MS Hamiltonian have clear physical significance, \textit{i.e.}, they describe the magnetic properties of the constituent ions and the coupling between them. Moreover, many of these parameters can be independently verified through measurements of related compounds~\cite{Yang2005InorgChem}. In contrast, the parameters deduced on the basis of a GSA are purely phenomenological. For example, comparisons between the two models have shown that higher order anisotropies in the GSA arise from the interplay between the local 2\textsuperscript{nd}-order single-ion anisotropy and the magnetic interactions between the ions, leading to mixing of excited spin states into the ground spin multiplet \cite{Wilson2006PRB,Hill2010Dalton,Maurice2010PRB}. In other words, the phenomenological $p\geq4$ zfs parameters are a direct manifestation of physics that goes beyond the GSA. The following section deals with this issue in detail, with a focus on the correlation between QTM behavior and the structural symmetries of real molecules.

\section{{Quantum Tunneling of Magnetization} in High-Symmetry {Mn$_3$} Single-Molecule Magnets}
\label{sec:Mn3}

The first clear observation of QTM selection rules, \textit{i.e.}, a complete absence of symmetry forbidden resonances \cite{Henderson2009PRL}, was reported for the trigonal SMM {[NE$_4$]$_3$[Mn$_3$Zn$_2$(salox)$_3$O(N$_3$)$_6$Cl$_2$]} (henceforth Mn$_3$) \cite{Feng2008InorgChem,Feng2009InorgChem}. This section focuses on QTM in SMMs with {trigonal symmetry}, emphasizing (i) symmetry-enforced {selection rules} that allow quantum relaxation in $k$-odd resonances (Sec.\,\ref{sec:SSRMn3}), (ii) the role of disorder (Sec.\,\ref{sec:disorderMn3}), and (iii) the microscopic origin of the $B_4^3\hat{O}_4^3$ GSA interaction which is predicted to give rise to unusual BPI patterns (Sec.\,\ref{sec:BPIMn3}).

\subsection{The Mn$_3$ Single-Molecule Magnet}
\label{sec:Mn3SMM}

\begin{figure}[t]
\centering
\includegraphics*[width=.7\textwidth]{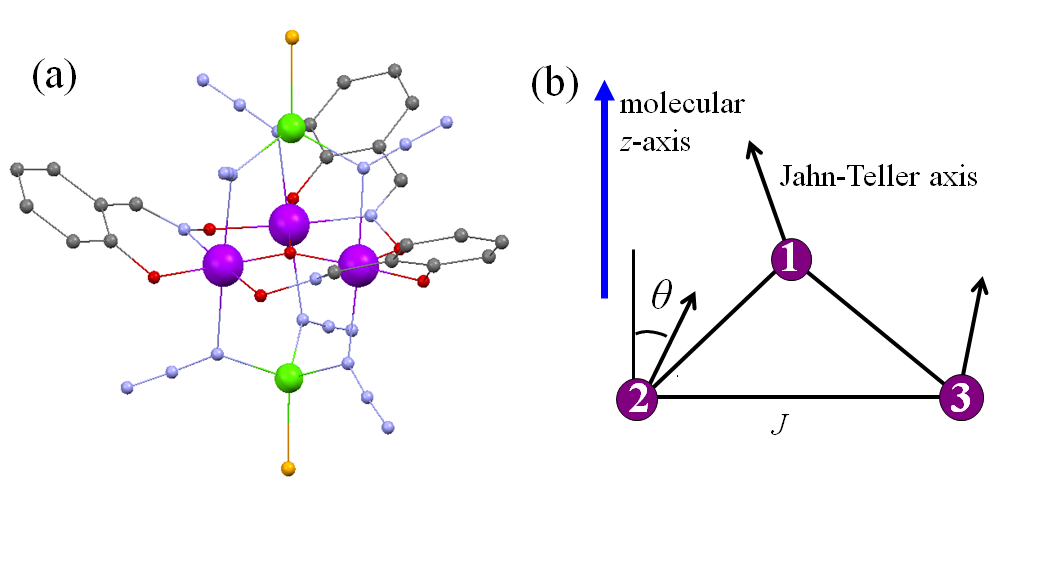}
\caption[]{The molecular structure (a) and schematic representation of the magnetic core (b) of the Mn$_3$ SMM. Color code: Mn = purple, Zn = green, O = red, N = blue, C = black and Cl = dark gold. H-atoms have been omitted for clarity.}
\label{fig:Mn3Structure}
\end{figure}

Several Mn$_3$ SMMs are known to crystallize in the trigonal space group $R3c$ with racemic mixtures of $C_3$ symmetric chiral molecules \cite{Feng2008InorgChem,Feng2009InorgChem,Hill2010Dalton}. The structure of the [NE$_4$]$_3$[Mn$_3$Zn$_2$(salox)$_3$O(N$_3$)$_6$Cl$_2$] molecule is shown in Fig.\,\ref{fig:Mn3Structure}a. The magnetic core consists of three ferromagnetically coupled Mn\textsuperscript{III} ($s =2$) ions, which form an equilateral triangle with two Zn\textsuperscript{II} ions located above and below the the Mn$_3$ plane, thus forming a trigonal bipyramidal structure. The $C_3$ axis of the molecule is perpendicular to the Mn$_3$ plane, while the local easy-axes of the individual spins are defined by the {Jahn-Teller (JT) elongation axes} of Mn\textsuperscript{III} ions, which are tilted slightly with respect to the $C_3$ axis. At low temperatures, the spin $S=6$ ground state multiplet can be described with the following GSA Hamiltonian:

\begin{equation}
\label{eqn:GSAMn3}
\hat{H}=D\hat{S}_z^2+B_4^0\hat{O}_4^0+B_4^3\hat{O}_4^3+B_6^6\hat{O}_6^6+\mu_B\vec{B}\cdot\buildrel{\lower3pt\hbox{$\scriptscriptstyle\leftrightarrow$}}\over{g}\cdot\vec{\hat{S}}
\end{equation}

\noindent{Due} to the {$C_3$ symmetry} of the molecule, the 2\textsuperscript{nd} order transverse anisotropy term, $E(\hat{S}_x^2-\hat{S}_y^2)$, is rigorously forbidden. Hence, the leading trigonal ($\hat{O}_4^3$) and hexagonal ($\hat{O}_6^6$) transverse zfs terms are instead included in Eqn.\,\ref{eqn:GSAMn3}.

Mn$_3$ is highly attractive in the context of understanding the origin of QTM at a microscopic level. The dimension of the MS Hamiltonian matrix for three $s = 2$ spins is just $[(2s + 1)^3]^2 = 125\times 125$. The $C_3$ symmetry reduces the number of interaction parameters to just a single exchange constant, $J$, and identical $d$ and $e$ values for each ion; it also guarantees identical $\theta_i$ Euler angles ($=15^\circ$) for the three spins, with $\phi_i = (i-1)\times120^\circ$. The remaining parameters have then been determined from fits to EPR and magnetization hysteresis measurements \cite{Feng2008InorgChem,Feng2009InorgChem,Henderson2009PRL,Hill2010Dalton}. Lastly, the structure contains no {solvent molecules}. This is rare among SMMs and removes a major source of disorder \cite{Redler2009PRB}. Consequently, exceptional spectroscopic data (QTM and EPR) are available against which one can test theoretical models. 

\subsection{{QTM Selection Rules} in Mn$_3$}
\label{sec:SSRMn3}

\begin{figure}[b]
\centering
\includegraphics*[width=.9\textwidth]{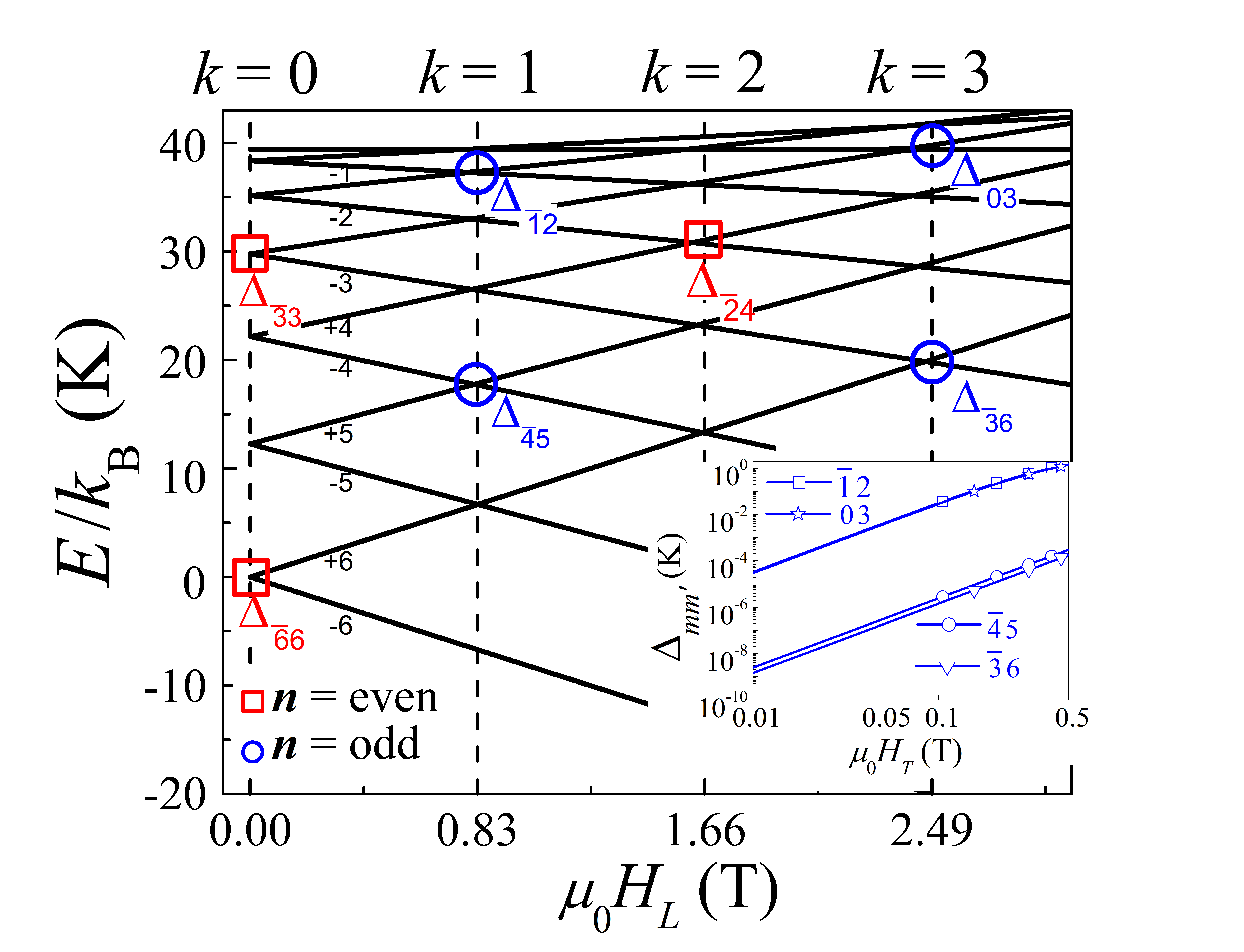}
\caption[]{Zeeman diagram for a spin $S = 6$ multiplet with easy-axis anisotropy ($D < 0$ in Eqn.\,\ref{eqn:GSAMn3}) and $H//z$. All possible non-zero tunneling gaps for $C_3$ symmetry are labeled according to the scheme discussed in the main text. The inset shows the $H_{\text{T}}$ dependence of the odd-$n$ tunneling gaps.}
\label{fig:Mn3Zeeman}
\end{figure}

For large spin systems, the effects of $q\not= 0$ zfs terms typically manifest themselves at energy scales that are orders of magnitude smaller than those of the axial ($q=0$) terms. One must therefore focus on {avoided level crossings}, where the {tunneling gaps} are governed by the transverse terms in Eqn.\,\ref{eqn:GSAMn3}. Fig.\,\ref{fig:Mn3Zeeman} displays the {Zeeman diagram} corresponding to the nominal spin $S=6$ ground state multiplet of the Mn$_3$ molecule. Due to symmetry restrictions ($q=3n$ for $C_3$ symmetry, where $n$ is an integer), non-zero tunneling gaps are limited to level crossings with $\textrm{\textDelta} m_s=3n$, where $m_s$ is the projection of the total spin onto the molecular $C_3$ (\textit{z}-) axis. All such gaps, $\text{\textDelta}_{m_s,m_s'}$, have been labeled in Fig.\,\ref{fig:Mn3Zeeman} for {QTM resonances} $k\le 3$, where $k$ ($=m_s+m_s'$) denotes an avoided crossing between pairs of levels with spin projections $m_s$ and $m_s'$ ($\bar{m}_s$ denotes $-|m_s|$).

By performing a mapping of the energy diagram obtained via exact diagonalization of Eqn.\,\ref{eqn:MS} onto that of the GSA Hamiltonian (Eqn.\,\ref{eqn:GSAMn3}) one can obtain microscopic insights into the emergence of $p\geq4$ transverse terms in the latter approximation. Published zfs parameters were employed for simulations involving Eqn.\,\ref{eqn:MS}, \textit{i.e.}, $d = -4.2$ K and $e = 0.9$ K \cite{Henderson2009PRL}. An isotropic exchange constant $J$ (= 10 K) was employed, set to a value that is artificially high in order to isolate the ground state from excited multiplets, thereby simplifying analysis of higher-lying QTM gaps (see Fig.\,\ref{fig:Mn3Zeeman}). The Euler angles were set to $\phi_1 = 0$, $\phi_2=120^{\circ}$ and $\phi_3=240^{\circ}$ (all $\psi_i=0$) to preserve $C_3$ symmetry, while $\theta_i$ ($=\theta$) was allowed to vary in order to examine its influence on QTM selection rules.

The situation in which the JT axes of the three Mn\textsuperscript{III} ions are parallel to the $C_3$ axis is first considered, \textit{i.e.}, $\theta =0$. The top section of Table.\,\ref{Table:Mn3mapping} lists the magnitudes of even-$n$ QTM gaps involving pairs of levels with $\text{\textDelta}m_s=3n$, deduced via diagonalization of Eqn.\,\ref{eqn:MS} in the absence of a transverse field, $H_{\text{T}}$~$(\perp z)$. The odd-$n$, $H_{\text{T}}=0$ gaps are identically zero, as can be seen from their dependence on $H_{\text{T}}$ (Fig.\,\ref{fig:Mn3Zeeman} inset): the power-law behavior indicates no contribution from zfs interactions (at $H_{\text{T}}=0$). Consequently, one expects only even-$n$ zfs terms of the form $B_{p}^{3n}\hat{O}_p^{3n}$ in the GSA: those satisfying this requirement have six-fold rotational symmetry about the $C_3$ axis, \textit{i.e.}, a higher symmetry than the real molecule (further explanation is given below). For comparison, these QTM gaps are simulated employing Eqn.\,\ref{eqn:GSAMn3} with $B_4^3 = 0$ and $B_6^6=4.3\times 10^{-7}$ K. As seen in Table\,\ref{Table:Mn3mapping}, an excellent overall agreement between the two models is obtained. Small differences may be attributed to higher-order six-fold terms such as $B_8^6\hat{O}_8^6$, $B_{10}^6\hat{O}_{10}^6$, \textit{etc.}, which have been neglected in this analysis.

A more realistic situation involves a tilting of the JT axes away from the $C_3$ axis by $\theta=8.5^{\circ}$, as is the case for Mn$_3$ \cite{Feng2009InorgChem}. Both even- and odd-$n$, $H_{\text{T}} = 0$ QTM gaps are generated in this situation, \textit{i.e.}, $k$-odd QTM resonances become allowed. This may be understood within the framework of the GSA as being due to the emergence of zfs interactions possessing three-fold rotational symmetry about the molecular $C_3$ axis, \textit{i.e.}, $B_p^{3n}\hat{O}_p^{3n}$ with $n=1$ and $p\geq4$; the leading such term is $B_4^3\hat{O}_4^3$. Table\,\ref{Table:Mn3mapping} lists the QTM gaps evaluated via diagonalization of Eqn.\,\ref{eqn:GSAMn3} using $B_6^6=4.3\times 10^{-7}$ K and $B_4^3=4.77\times 10^{-4}$~K. Excellent agreement is once again achieved between the GSA and MS Hamiltonians. Minor deviations may, in principle, be corrected by introducing higher-order transverse terms such as $B_6^3\hat{O}_6^3$.

\begin{figure}[t]
\centering
\includegraphics*[width=.7\textwidth]{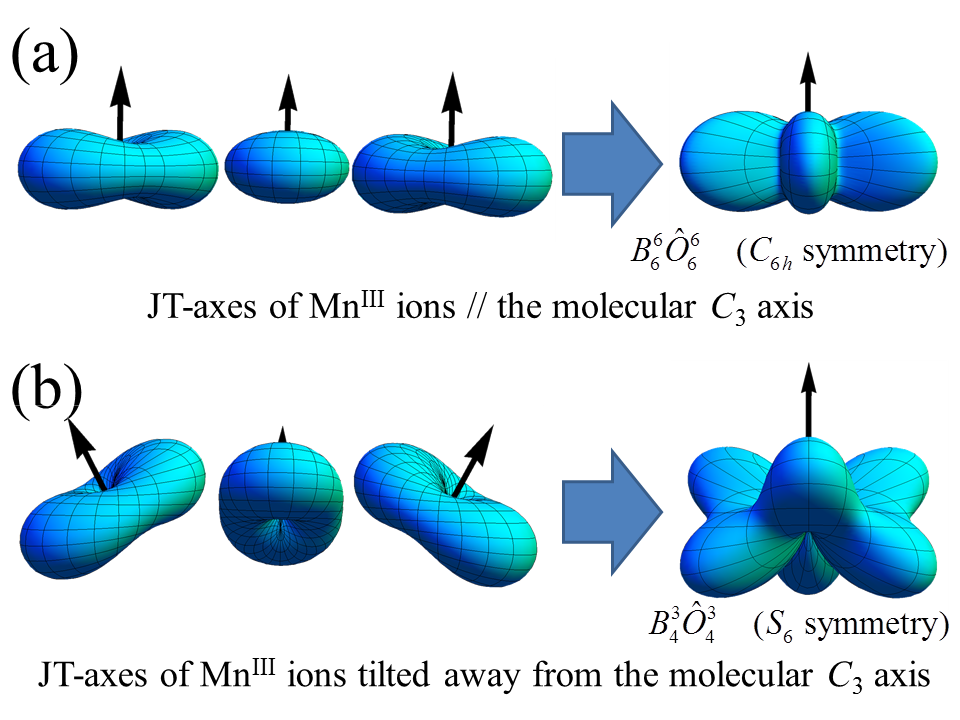}
\caption[]{The influence of the orientations of the JT-axes of Mn\textsuperscript{III} ions on the zero-field magneto symmetry of the Mn$_3$ SMM. In (a), the JT-axes of the Mn\textsuperscript{III} ions (left) are parallel to the molecule $C_3$ axis; consequently, the resultant Hamiltonian of the molecule (right) possesses $C_{6h}$ symmetry. In (b), the JT-axes of the Mn\textsuperscript{III} ions (left) are tilted away from the molecular $C_3$ axis; consequently, the resultant Hamiltonian of the molecule possesses $S_{6}$ symmetry}
\label{fig:Mn3MStoGSA}
\end{figure}

The emergence of the $B_4^3\hat{O}_4^3$ interaction in the GSA description of Mn$_3$ clearly indicates a lowering of the symmetry of the spin Hamiltonian upon tilting the JT axes. To understand this one must consider both the symmetry of the molecule and the intrinsic symmetry of the zfs tensors of the individual ions. Considering only 2\textsuperscript{nd} order SO anisotropy, the Hamiltonian of a single Mn\textsuperscript{III} ion possesses $D_{2h}$ symmetry (as noted in Sec.\,\ref{sec:GSA}), with three mutually orthogonal $C_2$ axes. When the JT axes are parallel ($\theta=0$), the local $z$-axis of each Mn\textsuperscript{III} center coincides with the molecular $C_3$ axis. The resultant Hamiltonian should then possess $C_3\times C_2\times C_i=C_{6h}$ symmetry (see Fig.\,\ref{fig:Mn3MStoGSA}a), requiring $B_4^3=0$; the additional $C_i$ symmetry arises from the {time-reversal invariance} of the SO interaction. In contrast, when the JT axes are tilted, the $C_2$ and $C_3$ axes do not coincide. In addition, the \textit{xy}-mirror symmetry of the molecule is broken, as is that of the spin Hamiltonian. The rotational symmetry then reduces to three-fold and, hence, $B_4^3\hat{O}_4^3$ is allowed; the symmetry in this case is $C_3 \times C_i = S_6$ (Fig.\,\ref{fig:Mn3MStoGSA}b).

It is possible to reinforce the preceding discussion via {group theoretic arguments}, without the need to write down an exact expression for the Hamiltonian. When the external magnetic field is applied parallel to the molecular $C_3$-axis, the $C_{6h}$ symmetry reduces to $C_6$, and the 13 basis functions of the $S=6$ ground state fall into six distinct one-dimensional {irreducible representations} \cite{Koster1963MITpress}. These functions can be grouped according to their behavior under a $C_6$ rotation: $\left|-6\right\rangle ,\left|0\right\rangle ,\left|+6\right\rangle \in \Gamma_1$; $\left|-2\right\rangle ,\left|+4\right\rangle \in \Gamma_2$;  $\left|+2\right\rangle ,\left|-4\right\rangle \in \Gamma_3$;  $\left|-3\right\rangle ,\left|+3\right\rangle \in \Gamma_4$;  $\left|+1\right\rangle ,\left|-5\right\rangle \in \Gamma_5$;  $\left|-1\right\rangle ,\left|+5\right\rangle \in \Gamma_6$, where $\Gamma_{1\ldots 6}$ are the six irreducible representations following the Bethe notation \cite{Koster1963MITpress}. Because the Hamiltonian operator belongs to the totally symmetric representation, $\left\langle m_s \right| \hat{H} \left| m_s' \right\rangle$ is non-zero only when $\left| m_s \right\rangle$ and $\left| m_s' \right\rangle$ belong to the same representation \cite{Cotton1990Wiley}. As can be seen, such states have $\text{\textDelta}m_s=3n$, with $n$ even, which is the criterion for state mixing in $C_6$ symmetry. When the symmetry of the Hamiltonian is reduced to $S_6$ ($C_3$ upon application of $H//z$) the basis functions may be grouped into three different irreducible representations: $\left|0\right\rangle ,\left|\pm 3\right\rangle ,\left|\pm 6\right\rangle \in \Gamma_1$;  $\left|+4\right\rangle ,\left|+1\right\rangle ,\left|-2\right\rangle,\left|-5\right\rangle \in \Gamma_2$; $\left|-4\right\rangle ,\left|-1\right\rangle ,\left|+2\right\rangle,\left|+5\right\rangle \in \Gamma_3$. Here, the selection rule for mixing is $\text{\textDelta}m_s=3n$, with $n$ being integer, again in agreement with the preceding calculations.

\begin{figure}[b]
\centering
\includegraphics*[width=.9\textwidth]{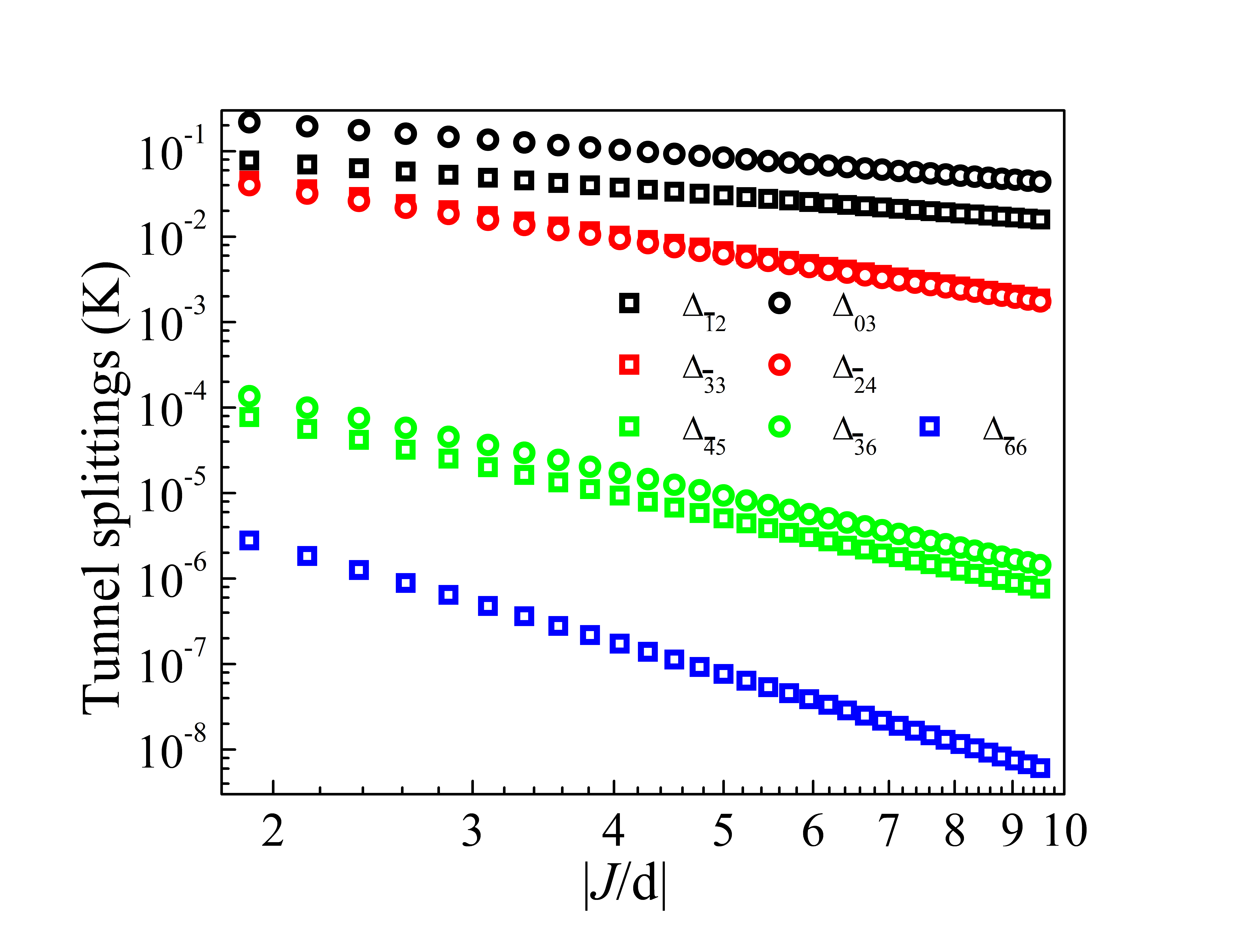}
\caption[]{Calculated QTM gaps for the Mn$_3$ SMM as a function of the coupling constant $J$. Simulations were performed with the JT-axes tilted $8.5^{\circ}$ away from the molecular $C_3$-axis. The QTM gaps associated with same $|\text{\textDelta}m|$ value are rendered in the same color. Note that the results are plotted on a logarithmic scale.}
\label{fig:Mn3Jdep}
\end{figure}

Before concluding this section, the influence of the {exchange coupling}, $J$, on the QTM observed in Mn$_3$ deserves further consideration. The $J$ dependence of higher-order ($p\geq 4$) coefficients in the GSA has been discussed previously for several other high-symmetry SMMs \cite{Hill2010Dalton,Wilson2006PRB,Liviotti2002JACS,Barra2007JACS,Maurice2010PRB}. In these cases, the 2\textsuperscript{nd}-order transverse anisotropy ($q>0$) cancels exactly, emerging at higher orders as a consequence of the mixing of spin states. This is illustrated for Mn$_3$ in Fig.\,\ref{fig:Mn3Jdep}, which plots the power law dependence of several QTM gaps as a function of the ratio of $J/d$; the single-ion zfs parameters given above were employed in these calculations. It is found that the QTM gaps are proportional to $|J|^{-n}$, \textit{i.e.}, $B_4^3 \propto |J|^{-1}$ and $B_6^6 \propto |J|^{-2}$ \cite{Hill2010Dalton}. Note that this implies a complete suppression of QTM in the {strong coupling} limit ($|J| \gg |d|$).

\subsection{The influence of {disorder} on QTM}
\label{sec:disorderMn3}

An important conclusion of the preceding analysis is the demonstration of the existence of $k$-odd QTM resonances, \textit{i.e}., a quite realistic parameterization of Eqn.\,\ref{eqn:MS} generates zfs terms in the GSA containing odd powers of $\hat{S}_+$ and $\hat{S}_-$. These ideas should apply quite generally. For example, the disorder potential associated with the distortion of a symmetric molecule likely contains zfs terms (\textit{e.g}. $\hat{O}_4^1$ or $\hat{O}_4^3$) that unfreeze $k$-odd QTM resonances (as explicitly demonstrated in Sec.\,\ref{sec:DisorderNi4}), contrary to the belief that odd QTM resonances \textit{cannot} be generated in this way~\cite{Slageren2009PRB}. However, it remains to be seen whether this can account for the absence of {selection rules} in SMMs such as {Mn$_{12}$}. We note that these arguments do not apply to zero-field ($k = 0$) QTM in half-integer spin systems, which is strictly forbidden according to {Kramers' theorem}~\cite{Wernsdorfer2002PRB}.

This revives a partly unresolved and somewhat controversial issue concerning the influence of disorder on the QTM characteristics of SMMs. Disorder became a focus of attention in some of the early spectroscopic investigations of the Mn$_{12}$-acetate and Fe$_8$Br SMMs, revealing significant distributions (or {strains}) in the measured GSA $D$ parameters \cite{Macca2001Poly,Park2001PRB,Hill2002PRB,Park2002PRBa}. Around the same time, Chudnovsky and Garanin argued that long-range strains nucleated by line dislocations could give rise to a broad distribution of transverse 2\textsuperscript{nd}-order anisotropies in otherwise high-symmetry crystals of SMMs such as {Mn$_{12}$-acetate}, \textit{i.e.}, a broad distribution (on a logarithmic scale) in $E$ centered about an average value of zero~\cite{Chudnovsky2001PRL,Chudnovsky:2002b}. Importantly, the Chudnovsky/Garanin theory pointed out that disorder would lead to local variations in molecular symmetry away from the ideal ($S_4$ for Mn$_{12}$-acetate), and that this could modify the selection rules governing QTM. This motivated intense efforts aimed at carefully studying QTM in Mn$_{12}$-acetate, including selective hole-burning experiments targeted at subsets of molecules belonging to different parts of the relaxation time distribution~\cite{Mertes2001PRL,delBarco2002EPL,delBarco2003PRL,delBarco:2005}. A breakthrough was achieved as a result of crystallographic investigations by Cornia \textit{et al.}~\cite{Cornia:2002}, that revealed a form of intrinsic disorder associated with the acetic acid solvent that co-crystallizes with the standard form of Mn$_{12}$-acetate (we refer the reader to Refs.~\cite{Cornia:2002,Hill2003PRL,delBarco2003PRL,Takahashi2004PRB,Bircher2004PRB,delBarco:2005} for detailed discussion). The acetic acid forms a hydrogen-bond to the Mn$_{12}$ core, resulting in a non-trivial distortion of the molecule. However, while each {solvent molecule} occupies a position between two Mn$_{12}$'s, it can only hydrogen-bond to one of them, with 50:50 probability. Hence, real Mn$_{12}$-acetate crystals contain a statistical distribution of several different solvent isomers, some of which maintain approximate four-fold symmetry, while more than $50\%$ have a lower (rhombic) symmetry~\cite{Cornia:2002}. {EPR}, {inelastic neutron scattering} and {magnetic hysteresis} measurements subsequently yielded excellent qualitative and quantitative agreement with the model proposed by Cornia, thus demonstrating for the first time that {solvent disorder} can have a profound influence on QTM relaxation \cite{Hill2003PRL,delBarco2003PRL,Takahashi2004PRB,Bircher2004PRB,delBarco:2005}.

Many more recent studies have reinforced the idea that solvent disorder can significantly influence QTM relaxation in SMMs. First of all, magnetization and EPR studies have shown that the anomalous distributions in zfs parameters found for Mn$_{12}$-acetate are absent in several newer high-symmetry ($S_4$) Mn$_{12}$ SMMs that do not suffer from the intrinsic solvent disorder (or for which the interaction between the solvent and the SMM core is far weaker than in the original acetate)~\cite{Hill2005Poly,Chakov2006JACS,Barra2007JACS,Redler2009PRB,Subedi2012PRB,Lamp2012IC}. Interestingly, the deliberate removal of solvent from the newer Mn$_{12}$'s (by pumping on the samples at room temperature) has been shown to accelerate the low temperature magnetization relaxation, without affecting the height of the classical relaxation barrier \cite{Redler2009PRB}. Meanwhile, EPR studies demonstrate that the solvent loss induces disorder that looks very similar to the intrinsic disorder in Mn$_{12}$-acetate~\cite{Redler2009PRB}. This again suggests that the induced (extrinsic) disorder causes the faster relaxation, presumably as a result of quantum tunneling processes. This leads to known sample handling problems, \textit{i.e.}, crystals containing volatile solvent (\textit{e.g.} Mn$_{12}$BrAc$\cdot$CH$_2$Cl$_2$) can change beyond recognition as far as their QTM and EPR characteristics are concerned if they are cooled under vacuum \cite{delBarco2004PRB,Redler2009PRB,Lamp2012IC}. 

The reason why the ideal Mn$_{12}$ SMM is so susceptible to disorder is because it has such a high symmetry; the nominally forbidden 2\textsuperscript{nd}-order transverse anisotropy rapidly reemerges upon the introduction of weak disorder, either through solvent loss or otherwise. This is not the case for lower symmetry molecules that already possess a 2\textsuperscript{nd}-order rhombic zfs interaction~\cite{Carbonera2010PRB}. This has caused some confusion in the literature. As an aside, we note that internal transverse dipolar/hyperfine fields can, in principle, also affect QTM selection rules in high-spin SMMs~\cite{Burzuri2009PRB}. Indeed, early work demonstrated that a combination of allowed transverse zfs interactions, together with transverse {dipolar/hyperfine fields}, may explain the observed absence of QTM selection rules in Mn$_{12}$-acetate in the {thermally activated regime} \cite{Fernandez:1998,Luis:1998}. However, this explanation fails in the {pure QTM regime}, where tunneling couples low lying spin states at a high order of perturbation theory ($\text{\textDelta}m_s >> 1$). More recent studies claim that differences in QTM relaxation observed for the various high-symmetry Mn$_{12}$ SMMs are due entirely to differences in the widths of the {dipolar field distributions}, which obviously depend on the crystal structures~\cite{Burzuri2009PRB}. However, the recent preparation of two versions of Mn$_{12}$-acetate that are identical in almost all respects (including the lattice constants), apart from the co-crystallizing solvent (acetic acid in one case, methanol in the other), seem to rule out this assertion~\cite{Redler2009PRB}. While dipolar fields undoubtedly play a crucial role in the collective QTM relaxation in SMM crystals~\cite{Stamp1998PRL}, the marked differences in relaxation rates found for the two Mn$_{12}$-acetates appear to be related to the disorder associated with the hydrogen bonding acetic acid solvent, which is not present in the methanol variant.

The large dimension of the Mn$_{12}$ MS Hamiltonian presents a considerable challenge in terms of gaining theoretical insights into the effects of disorder. However, there exist many smaller molecules with equally high symmetry which are, thus, more amenable to this type of study. Indeed, this issue is revisited in Sec.\,\ref{sec:DisorderNi4}, which deals explicitly with the distortion of a {Ni$_4$} SMM that possesses the same intrinsic $S_4$ symmetry as the ideal Mn$_{12}$'s~\cite{Wilson2006PRB}. Aside from the obvious computational advantages, several smaller SMMs are also known to crystallize with no lattice solvent molecules~\cite{Lawrence2008IC,Feng2008InorgChem,Hill2010Dalton}. More importantly, there exist families of low-nuclearity SMMs for which some members co-crystallize with solvent, while others do not. These include {Mn$_3$}, {Mn$_4$} and Ni$_4$, which represent the focus of the remainder of this chapter. The spectroscopic differences between solvated and solvent-free SMMs are quite dramatic. For instance, {$D$-strain} is almost absent in the latter, giving rise to remarkably sharp EPR spectra. This again implicates solvent molecules as a major source of disorder in SMM crystals. The key finding involved a solvent-free Mn$_3$ compound, which is the only SMM to display a complete absence of a symmetry-forbidden QTM resonance~\cite{Henderson2009PRL}. When combined with the observation of uniquely sharp EPR spectra~\cite{Feng2008InorgChem}, this result suggests that it is the absence of solvent that unmasks the intrinsic QTM {selection rules}, again hinting at the connection between solvent, disorder and the absence of QTM selection rules in other SMMs.

\begin{figure}[t]
\centering
\includegraphics*[width=1.0\textwidth]{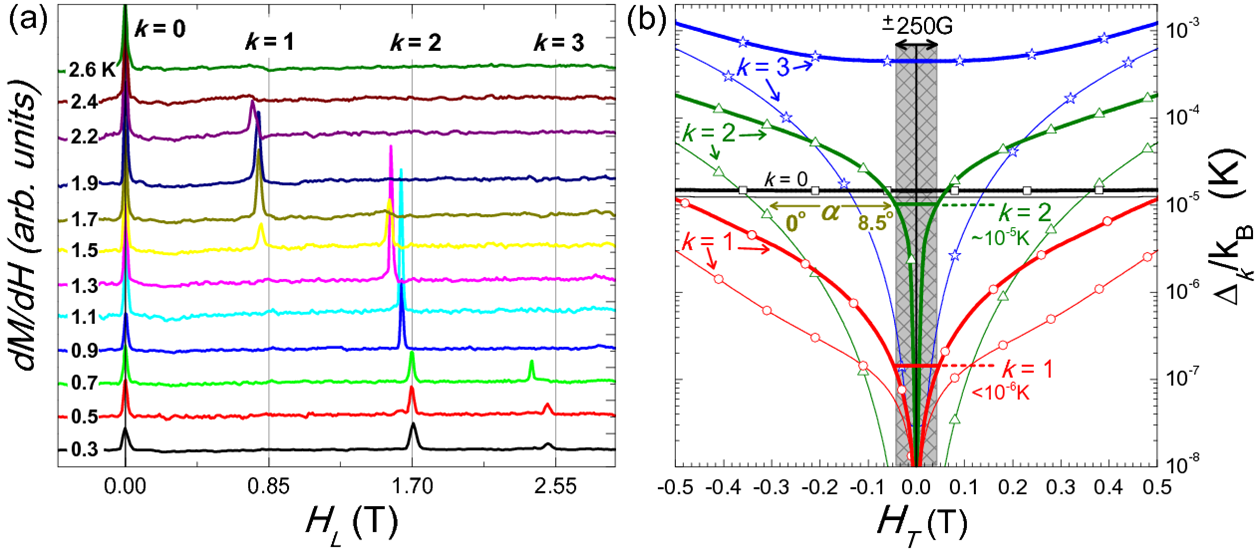}
\caption[]{(a) Field derivative of the magnetization curves obtained for a Mn$_3$ single crystal at different temperatures, with $B\parallel z$. (b) Ground-state tunnel splittings associated with resonances $k=0$ (black squares), $k=1$ (red circles), $k=2$ (green triangles), and $k=3$ (blue stars) as a function of the transverse field $H_{\text{T}}$, with the JT-axes aligned along the $C_3$ axis (thin lines) and tilted by 8.5 degrees away from the $C_3$ axis (thick lines). The strength of the dipolar magnetic field in the sample is represented by the central gray area, with the corresponding splitting values achieved for such dipolar field values for resonances $k=1$ and $k=2$ (dashed horizontal lines).}
\label{fig:Mn3PRL}
\end{figure}

Fig.\,\ref{fig:Mn3PRL}a shows derivatives of magnetization hysteresis curves for Mn$_3$, recorded at different temperatures from 0.3 K to 2.6 K, with $H\parallel z$. At low temperatures, the $k=1$ resonance is completely absent. It eventually appears for temperatures above 1.3 K as a result of a symmetry allowed thermally activated QTM process. As discussed above, the trigonal symmetry of this molecule enforces the $|\text{\textDelta}m|=3n$ selection rule when taking into account the 8.5 degree misalignment of the JT axes from the molecular $C_3$ axis ($S_6$ reduced to $C_3$ when a longitudinal field is applied). The effect can be seen in Fig.\,\ref{fig:Mn3PRL}b, which shows the {tunnel splittings} for the four lowest resonances $k=0-3$, calculated by diagonalization of the MS Hamiltonian of Eqn.\,\ref{eqn:MS} with the parameters given in Ref.\cite{Henderson2009PRL}. In the absence of a transverse field ($H_T=0$), the ground state tunnel splitting is always absent for resonances $k=1$ and $k=2$, while the degeneracy is only broken in resonances $k=0$ and $k=3$. Consequently, one expects steps in the hysteresis curves (peaks in the derivatives) appearing only at $k=0$ and $k=3$. The absence of the $k=1$ resonance at low temperatures constitutes direct evidence for the expected QTM selection rule, an observation made possible because of the highly ordered solvent-free crystal structure. Following the same reasoning, resonance $k=2$ should also be absent at low temperatures, since the ground tunnel splitting couples spin states differing by $|\text{\textDelta}m|\neq 3n$. However, as can be seen in Fig.\,\ref{fig:Mn3PRL}b, the $k=2$ splitting grows quickly as a function of the transverse field, reaching observable magnitudes for field values provided by internal {dipolar fields}. This result shows that internal Zeeman fields can indeed unfreeze some QTM resonances, but not all of them. Indeed, the ground state tunnel splitting associated with the $k=1$ resonance remains almost two orders of magnitude smaller than that of $k=2$ for the same transverse field. One would expect the influence of dipolar fields to diminish further still in the pure QTM regime for SMMs with larger $S$ values.

The Mn$_3$ SMM illustrates perfectly how crystalline disorder can mask the fundamental QTM behavior in SMMs; in this particular case, it is the absence of disorder that unmasks intrinsic symmetry-enforced quantum properties. This, in turn, allows fundamental insights into the influence of the internal molecular degrees of freedom on the QTM phenomenon. The low-nuclearity of the Mn$_3$ SMM proved particularly helpful by making this a computationally tractable problem. The following section digs deeper into the unusual BPI patterns predicted for trigonal SMMs.

\subsection{{Berry Phase Interference} in {Trigonal Symmetry}}
\label{sec:BPIMn3}

This section focuses explicitly on the BPI patterns generated by the $\hat{O}_4^3$ operator. In contrast to all of the even-$q$ GSA terms, the \textit{xy}-plane does not correspond to a symmetry element for the odd-$q$ interactions, as discussed in Sec.\,\ref{sec:HighOrder}. Hence, the $\hat{O}_4^3$ operator is expected to result in BPI patterns which have not been observed in previous studies of SMMs, essentially all of which possess even rotational symmetry with only even-$q$ zfs interactions \cite{Garg1993EPL,Park2002PRB,Li2011PRB}.

\begin{figure}[b]
\centering
\includegraphics*[width=.7\textwidth]{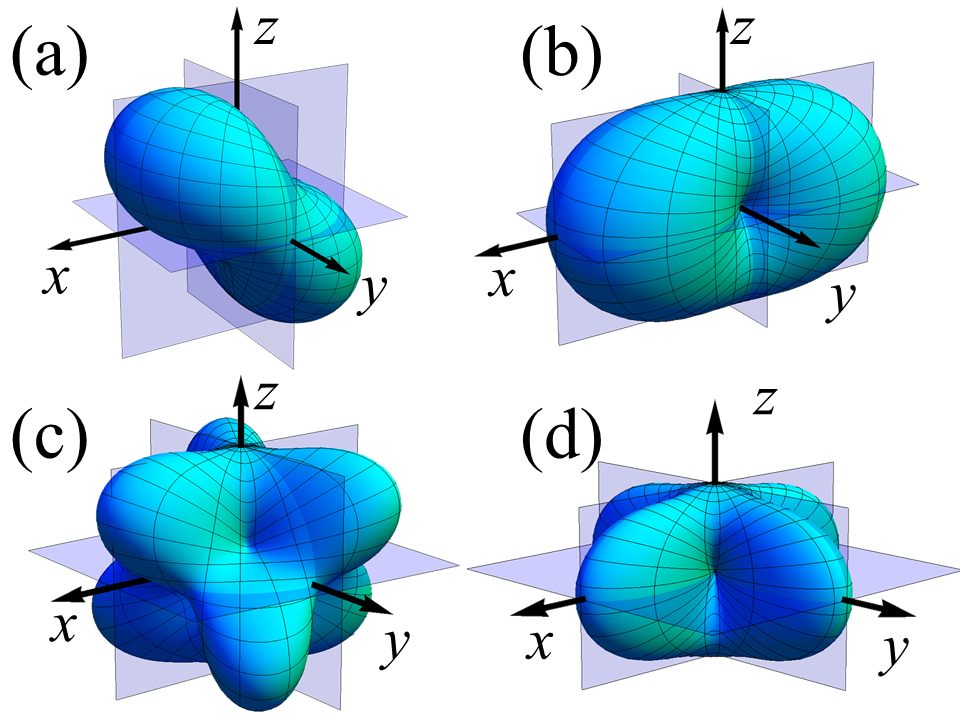}
\caption[]{Potential energy surfaces corresponding to the fourth order Stevens operators $\hat{O}_4^1$ (a), $\hat{O}_4^2$ (b), $\hat{O}_4^3$ (c) and $\hat{O}_4^4$ (d). As can be seen, the $q$-even operators include the \textit{xy}-plane as an extra symmetry element while the $q$-odd operators include improper rotations (see main text).}
\centering
\label{fig:GSAHighOrder}
\end{figure}

\begin{figure}[b]
\centering
\includegraphics*[width=1.0\textwidth]{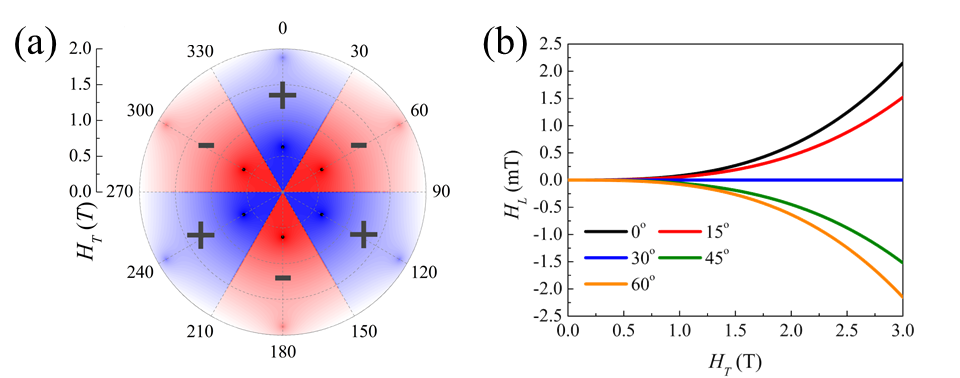}
\caption[]{(a) The calculated BPI patterns associated with the ground state $k = 0$ QTM resonance for the Mn$_3$ SMM: the color contour plot shows \textDelta$_{\bar{6},6}$ as function of $H_{\text{T}}$ (with $B_6^6$ set to zero); a compensating $H_{\text{L}}$ field is required that alternates between positive (red) and negative (blue) values. (b) The compensating $H_{\text{L}}$ field for \textDelta$_{\bar{6},6}$, as a function of the magnitude of $H_{\text{T}}$; note the curvature (except for the 30$^{\circ}$ trace, for which $H_{\text{L}}=0$).}
\label{fig:Mn3BPI}
\end{figure}

The influence of $B_4^3\hat{O}_4^3$ is quite fascinating. In order to simplify discussion, Fig.\,\ref{fig:Mn3BPI} was generated with $B_6^6=0$; details of the calculations can be found in Ref \cite{Liu2012PRB}. The \textDelta$_{\bar{6},6}$ ($k=0$) QTM resonance exhibits the most intriguing new features. One might expect a six-fold behavior due to the intrinsic $C_i$ symmetry of the Hamiltonian, \textit{i.e.}, the spectrum should be invariant under inversion of $H_{\text{T}}$. However, this assumes that $H_{\text{L}}=0$. In fact, application of a {transverse field} causes a shift of the \textDelta$_{\bar{6},6}$ minimum away from $H_{\text{L}}=0$, as illustrated in Fig.\,\ref{fig:Mn3BPI}b, \textit{i.e.}, the $k=0$ QTM resonance shifts away from $H_{\text{L}}=0$ in the presence of a finite transverse field. The resultant transverse-field BPI patterns appear to exhibit a hexagonal symmetry in Fig.\,\ref{fig:Mn3BPI}a. However, the color coding represents the polarity of the required compensating longitudinal field, $H_{\text{L}}$. Thus, on the basis of the sign of $H_{\text{L}}$, one sees that the BPI minima in fact exhibit a three-fold rotational symmetry, which is consistent with the symmetry of the $B_4^3\hat{O}_4^3$ interaction. One way to interpret this result is to view the $\hat{O}_4^3$ operator as generating an effective internal longitudinal field, $H^{*}_{\text{L}}$, under the action of an applied transverse field; $H^{*}_{\text{L}}$ is then responsible for the shift of the $k = 0$ resonance from $H_{\text{L}}=0$. This can be seen from the expression of the $\hat{O}_4^3=\frac{1}{2}[\hat{S}_z,\hat{S}_-^3+\hat{S}_+^3]$ operator, which, unlike the $q$-even operators, contains an odd power of $\hat{S}_z$, akin to the Zeeman interaction with $H\parallel z$. An alternative view may be derived from the $S_6$ surface depicted in Fig.\,\ref{fig:GSAHighOrder}c, where one sees that the hard/medium directions do not lie within the $xy$-plane, contrary to the case for the $q$-even operators. In other words, the classical hard plane is not flat, but corrugated with a 120$^{\circ}$ periodicity. Consequently, application of a longitudinal field is required in order to insure that the total applied field lies within the hard plane when rotating $H_{\text{T}}$. Note that the predicted BPI patterns nevertheless exhibit the required $C_i$ symmetry, \textit{i.e.}, they are invariant with respect to inversion of the total field.

Fig.\,\ref{fig:Mn3BPI}b plots the shift of the $k=0$ resonance (\textDelta$_{\bar{6},6}$ minimum) away from $H_{\text{L}}=0$ upon applying a transverse field, $H_{\text{T}}$, for several orientations within the $xy$-plane. The shift is positive for 0$^{\circ}$ and 15$^{\circ}$, and negative for 45$^{\circ}$ and 60$^{\circ}$, with no shift at 30$^{\circ}$ (\textit{i.e.} the 30$^{\circ}$ resonance occurs at $H_{\text{L}}=0$). In other words, the quantum molecular hard plane is not flat, but rather corrugated, with a 120$^{\circ}$ periodicity. This is consistent with the classical energy surface shown in Fig.\,\ref{fig:GSAHighOrder}c. It is also notable that the $H_{\text{L}}$ shift displays a non-linear dependence on $H_{\text{T}}$, which indicates that the exact locations of the hard directions depend on the magnitude of $H_{\text{T}}$. Finally, it should be emphasized that these phenomena, especially the shift of the $k=0$ resonance from $H_{\text{L}}=0$, cannot be generated by any of the even-$q$ operators \cite{Li2011PRB}, where the \textit{xy} plane necessarily corresponds to the hard plane because of the additional mirror symmetry about this plane (see the discussion in Sec.\,\ref{sec:HighOrder}).

\section{Quantum Tunneling of Magnetization in the High-Symmetry {Ni$_4$} Single-Molecule Magnet}
\label{sec:Ni4}

\subsection{The Ni$_4$ Single-Molecule Magnet}
\label{sec:Ni4SMM}

\begin{figure}[t]
\centering
\includegraphics*[width=.7\textwidth]{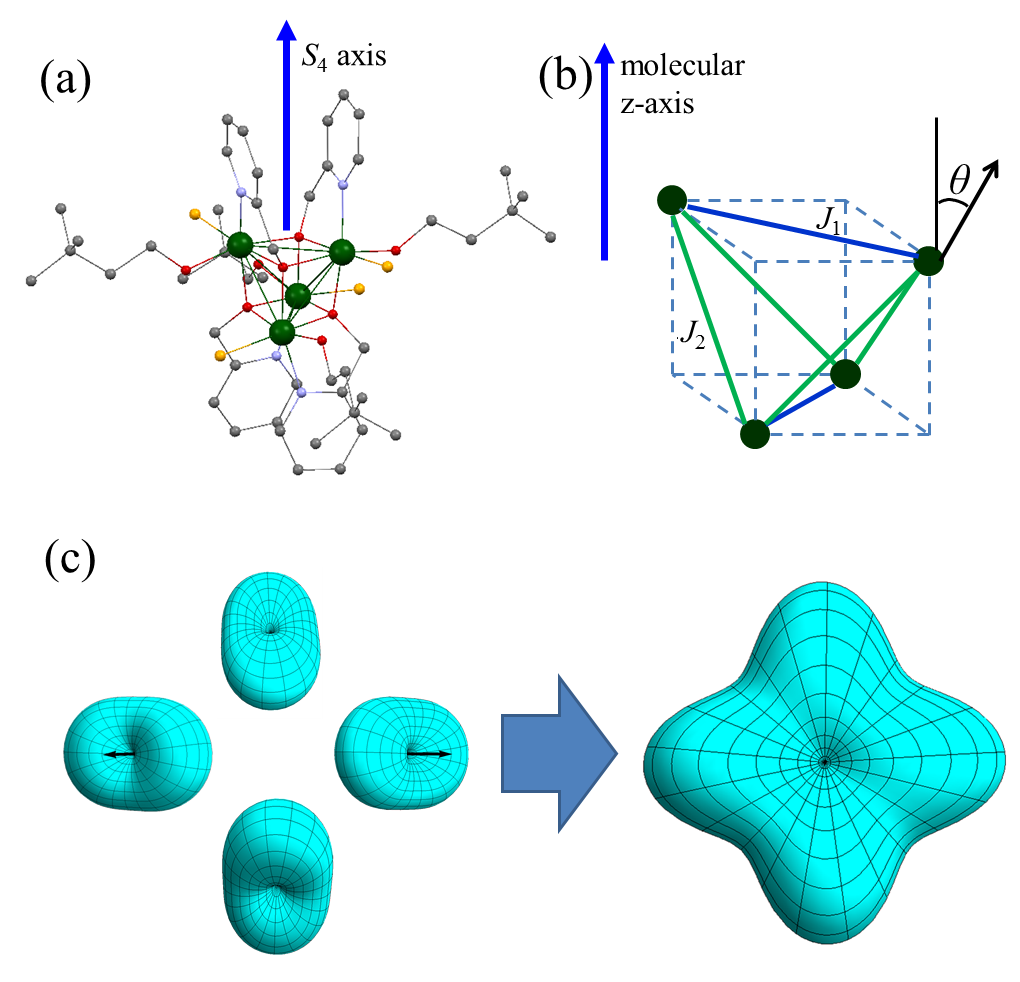}
\caption[]{The structure (a) and schematic representation of the magnetic core (b) of the Ni$_4$ SMM. Color code: Ni = olive, O = red, N = blue, C = black and Cl = dark gold. H-atoms have been omitted for clarity. (c) Representation of the zero-field magneto symmetry of the Ni$_4$ SMM resulting from the situation in which the 2\textsuperscript{nd}-order single-ion zfs tensors have their $C_2$ axes tilted away from the molecular $S_4$ axis. Once added, the time reversal symmetry of the SO interaction guarantees that the resultant zero-field Hamiltonian of the molecule possesses $C_{4h}$ symmetry (see text for details).}
\label{fig:Ni4Structure}
\end{figure}

The {[Ni(hmp)(dmb)Cl]$_4$} SMM (henceforth Ni$_4$) possesses $q$-even rotational symmetry \cite{Yang2003Polyhedron,Kirman2005JAP,Wilson2006PRB,Yang2006InorgChem,Lawrence2008IC}. The complex crystallizes in an $I41/a$ space group without any lattice solvent molecules. The structure of the Ni$_4$ molecule is shown in Fig.\,\ref{fig:Ni4Structure}a. The magnetic Ni$_4$O$_4$ core is a slightly distorted cube with the Ni$^{\text{II}}$ ions ($s=1$) located on opposite corners, as sketched in Fig.\,\ref{fig:Ni4Structure}b. The distorted cube retains {$S_4$ symmetry}, with the $S_4$-axis indicated in Fig.\,\ref{fig:Ni4Structure}a. The four Ni\textsuperscript{II} ions are ferromagnetically coupled, leading to a spin $S = 4$ molecular ground state. The Ni$_4$ SMM exhibits extremely fast zero-field QTM, which significantly reduces the effective relaxation barrier. Nevertheless, it does display a small magnetic hysteresis \cite{Yang2006InorgChem}. However, the {fast relaxation} unfortunately precludes the observation of $k>0$ QTM resonances. Nevertheless, a theoretical study of Ni$_4$ proves enlightening. The molecule can be described with the following {spin Hamiltonian}:

\begin{equation}
\label{eqn:Ni4MS}
{\hat H_{{\rm{zfs}}}} = \sum\limits_{i = 1}^4 {{{\hat s}_i} \cdot {{({{\mathord{\buildrel{\lower3pt\hbox{$\scriptscriptstyle\leftrightarrow$}} 
\over n} }^T})}^i} \cdot \mathord{\buildrel{\lower3pt\hbox{$\scriptscriptstyle\leftrightarrow$}} 
\over d}  \cdot {{(\mathord{\buildrel{\lower3pt\hbox{$\scriptscriptstyle\leftrightarrow$}} 
\over n} )}^i} \cdot {{\hat s}_i}}  + \sum\limits_{i < j} {{{\hat s}_i} \cdot {{\mathord{\buildrel{\lower3pt\hbox{$\scriptscriptstyle\leftrightarrow$}} 
\over J} }_{ij}} \cdot {{\hat s}_i}} .
\end{equation}

\noindent{This} Hamiltonian differs from Eqn.\,\ref{eqn:MS} in that the individual rotation matrices, $\buildrel{\lower3pt\hbox{$\scriptscriptstyle\leftrightarrow$}} \over R_i$, are replaced by a single matrix, $\buildrel{\lower3pt\hbox{$\scriptscriptstyle\leftrightarrow$}} \over n$, which explicitly takes the rotational symmetry of the molecule into account, including cases involving improper rotations. The zero-field anisotropy can then be parameterized by a single $\buildrel{\lower3pt\hbox{$\scriptscriptstyle\leftrightarrow$}} \over d$-tensor (corresponding to one of the ions), specified with respect to the molecular coordinate frame. If the local coordinate frames of the individual ions are related by a series of proper rotations ($C_2$, $C_4$, \textit{etc}.) within the molecular coordinate frame, then $\buildrel{\lower3pt\hbox{$\scriptscriptstyle\leftrightarrow$}} \over n$ may be replaced by a single rotation matrix $\buildrel{\lower3pt\hbox{$\scriptscriptstyle\leftrightarrow$}} \over R$ (corresponding, \textit{e.g.}, to a $90^\circ$ rotation about $z$ for a molecule with $C_4$ symmetry). On the other hand, if the local coordinate frame of the \textit{i}\textsuperscript{th} ion is related to the molecular frame by an {improper rotation} ($S_4$ or $S^3_4$), then $\buildrel{\lower3pt\hbox{$\scriptscriptstyle\leftrightarrow$}} \over n=\sigma\buildrel{\lower3pt\hbox{$\scriptscriptstyle\leftrightarrow$}} \over R$, where $\sigma$ represents a reflection in the plane perpendicular to the $S_4$ axis. Note that, for $S_4$ symmetry, $\buildrel{\lower3pt\hbox{$\scriptscriptstyle\leftrightarrow$}} \over n^2$ is equivalent to a $C_2$ rotation, and $\buildrel{\lower3pt\hbox{$\scriptscriptstyle\leftrightarrow$}} \over n^4$ is equivalent to the identity matrix.

\begin{figure}[b]
\centering
\includegraphics*[width=.9\textwidth]{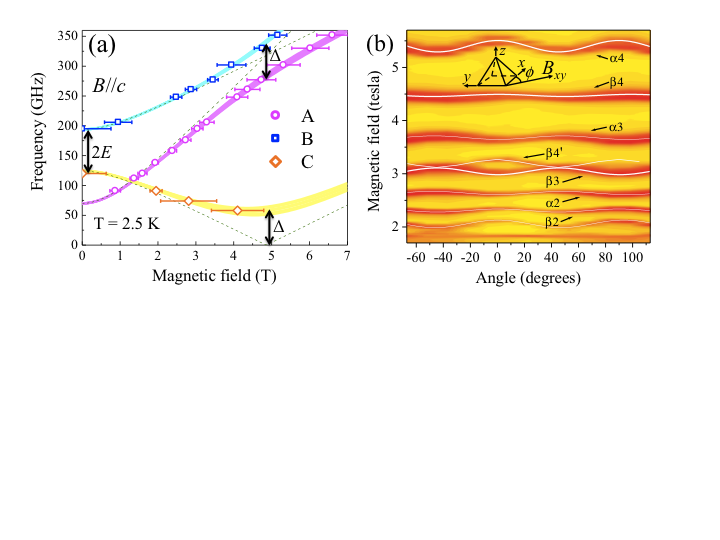}
\caption[]{(a) Frequency dependence of the positions (in field) of the three EPR transitions associated with the isolated Ni\textsuperscript{II} triplet ($S=1$) state in a diluted single crystal of the compound [Zn$_{3.91}$Ni$_{0.09}$(hmp)$_4$(dmb)$_4$Cl$_4$] (see Ref.~\cite{Yang2005InorgChem} for assignments of the A, B and C peaks). The colored curves correspond to best fits to the data employing the following single-ion zfs parameters: $d = -5.30(5)$~cm$^{-1}$, $e = \pm1.20(2)$~cm$^{-1}$, $g_z=2.30(5)$, and a tilting of the local $z$-axes of $15^\circ$ away from the symmetry ($c$-) axis of the crystal. The energy splittings labeled \textDelta~provide a direct measure of the tilting of the local $\buildrel{\lower3pt\hbox{$\scriptscriptstyle\leftrightarrow$}} \over d$ tensors; the dashed curves correspond to the non-tilted case, for which these splittings are zero. The widths of the colored curves reflect the uncertainty in the orientations of the local $x$- and $y$-axes, which were subsequently deduced from two axis rotation studies \cite{Yang2005InorgChem}. (b) 2D color map of the EPR absorption intensity as a fucntion of the magnetic field strength and its orientation within the hard plane of a single crystals of the $S=4$ SMM [Ni$_4$(hmp)(dmb)Cl]$_4$ (see Ref.~\cite{Lawrence2009PCCP} for explanation of the peak labeling). Superimposed on the absorption maxima (darker red regions) are fits (white curves) to the data that involve just a single adjustable zfs parameter, $B_4^4 = 4\times10^{-4}$~cm$^{-1}$ (over and above those deduced from easy-axis measurements~\cite{Wilson2006PRB}).}
\label{fig:Ni4Data}
\end{figure}

The Ni$_4$ SMM is a particularly ideal platform for comparison with Mn$_3$. The molecule possesses a well separated $S = 4$ ground state with the $S = 3$ excited spin multiplets located roughly 30~K above in energy. The $3 \times 3$ Hamiltonian matrix associated with a single Ni\textsuperscript{II} ion contains only two 2\textsuperscript{nd}-order zfs parameters, $d$ and $e$, \textit{i.e.}, higher order single-ion anisotropies ($p \geq 4$) are strictly forbidden. The zfs of the individual Ni\textsuperscript{II} ions, as well as their orientations, have been directly measured through {EPR} studies on an isostructural diluted Zn$_{4-x}$Ni$_x$ compound (see  Fig.\,\ref{fig:Ni4Data}(a) and Ref.~\cite{Yang2005InorgChem}). Due to the restriction of $S_4$ symmetry, only two independent Heisenberg interaction parameters, $J_1$ and $J_2$, are allowed; these interactions can be determined by dc susceptibility measurements \cite{Yang2006InorgChem}. Therefore, all of the parameters in Eqn.\,\ref{eqn:Ni4MS} are known independently. Meanwhile, the molecule possesses the same $S_4$ symmetry as Mn$_{12}$, which prohibits the rhombic anisotropy term in the GSA Hamiltonian. The high symmetry of the molecule has been confirmed by single-crystal EPR measurements, where exceptionally sharp resonances are again observed, with a four-fold modulation pattern upon rotating $H_{\text{T}}$ (see  Fig.\,\ref{fig:Ni4Data}(b) and Ref.~\cite{Lawrence2009PCCP}). This clearly illustrates the presence of a 4\textsuperscript{th}- (or higher-) order transverse GSA anisotropy which is responsible for the fast QTM.

\subsection{{Quantum Tunneling of Magnetization} in the {Ni$_4$} SMM}
\label{sec:Ni4QTM}

\begin{figure}[t]
\centering
\includegraphics*[width=.7\textwidth]{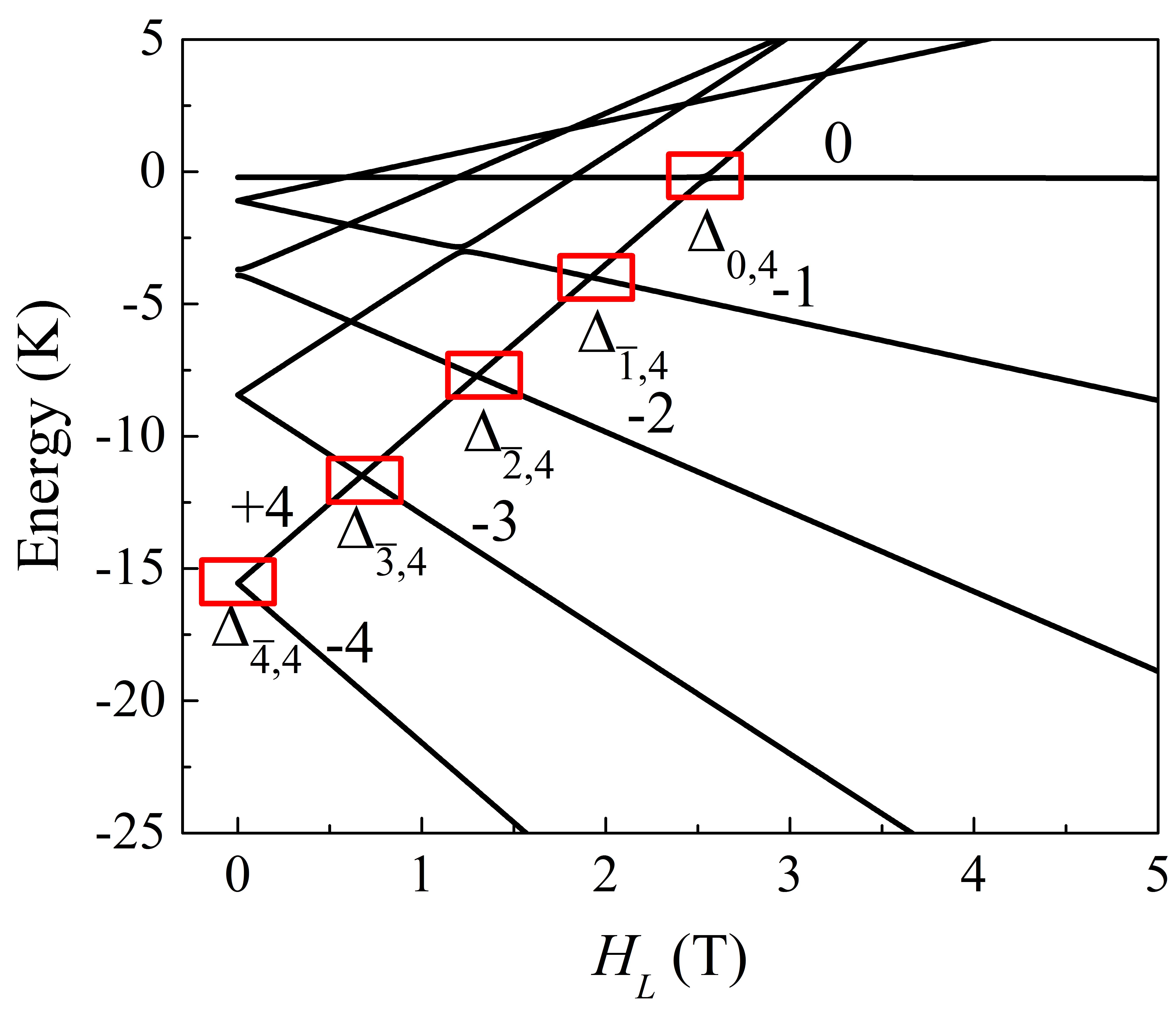}
\caption[]{Zeeman diagram for the ground state $S=4$ multiplet associated with the Ni$_4$ SMM, simulated employing Eqn.\,\ref{eqn:Ni4MS}. The $k$ = 0 to 4 ground state QTM splittings are labeled in the figure.}
\label{fig:Ni4Zeeman}
\end{figure}

In analogy with Mn$_3$, the transverse GSA anisotropy in Ni$_4$ is assessed by calculating the QTM gaps, focusing on the $k=0,1\ldots 4$ ground state resonances, as shown in Fig.\,\ref{fig:Ni4Zeeman}; the $|\text{\textDelta}m|$ values associated with these resonances equal $8,7\ldots 4$, respectively. The simulations were performed using the published zfs parameters $d$ = -7.6 K, $e$ = 1.73 K and $J_1$ = $J_2$ = -10 K \cite{Wilson2006PRB,Yang2005InorgChem,Kirman2005JAP}. Previous EPR studies also show that the local easy-axes of the Ni\textsuperscript{II} ions are tilted away from the molecular \textit{z}-axis by $\theta = 15^{\circ}$ (See Fig.\,\ref{fig:Ni4Structure}c). However, the $\theta = 0^{\circ}$ case is also examined in order to further explore the influence of easy-axis tilting on the symmetry of the molecular Hamiltonian.

\begin{figure}[t]
\centering
\includegraphics*[width=.7\textwidth]{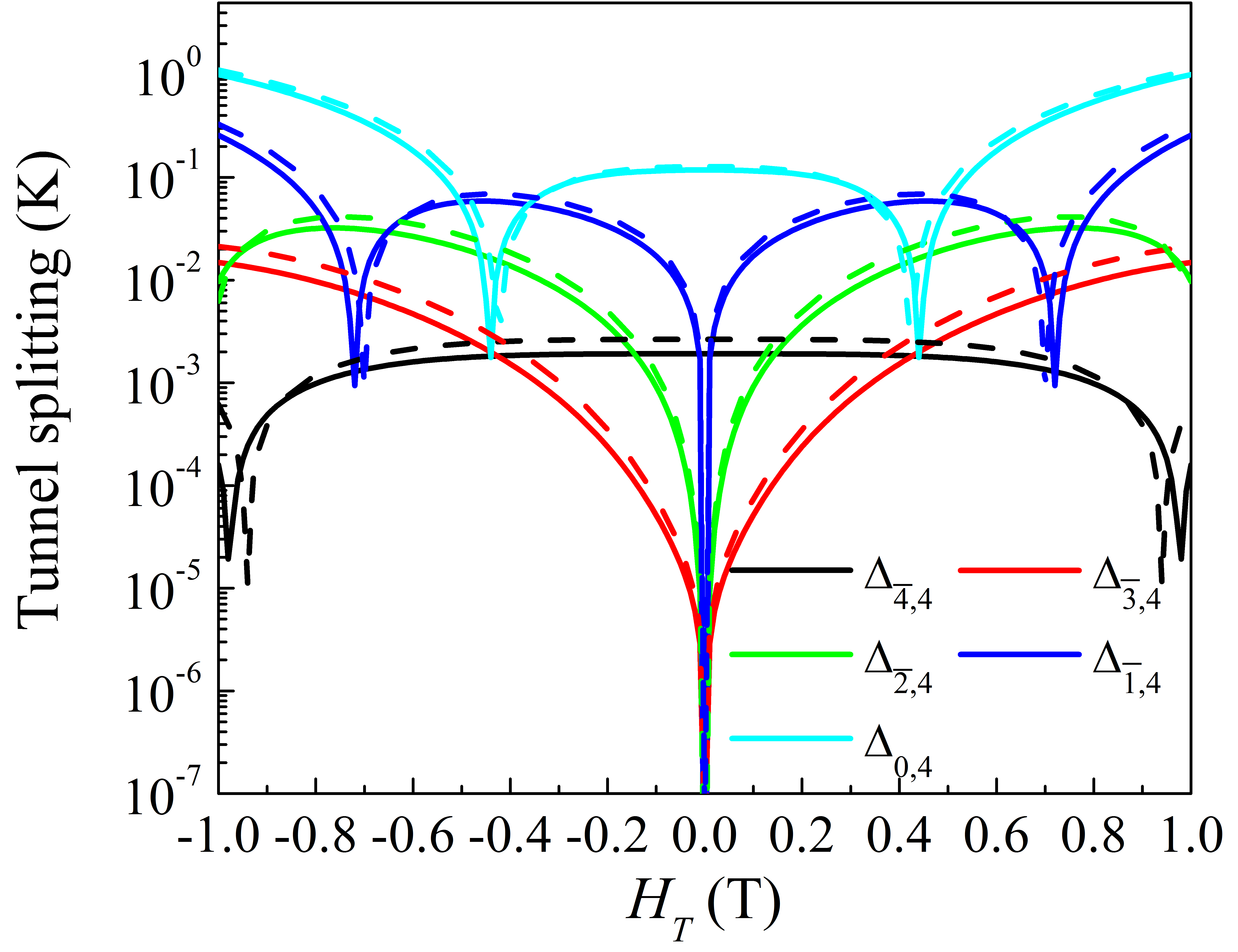}
\caption[]{The ground state QTM gaps for the Ni$_4$ SMM as a function of $H_{\text{T}}$. The simulations were performed employing Eqn.\,\ref{eqn:Ni4MS} with the parameters given in the main text. The solid lines were generated with $\theta = 0$ and the dash lines were generated with $\theta = 15^{\circ}$.}
\label{fig:Ni4Gap}
\end{figure}

Fig.\,\ref{fig:Ni4Gap} shows the ground state QTM gaps as function of transverse field ($H_{\text{T}}$), deduced via exact diagonalization of Eqn.\,\ref{eqn:Ni4MS}. As seen in the figure, \textDelta$_{\bar{4},4}$ ($k=0$) and \textDelta$_{0,4}$ ($k=4$) retain non-zero values in the absence of a transverse field, while all other {tunnel splittings} vanish at $H_{\text{T}}=0$. This result is not surprising based on the $S_4$ molecular symmetry, where only $|\text{\textDelta}m|=4n$ ($n$ is an integer) QTM resonances are allowed. However, unlike the Mn$_3$ SMM, the QTM selection rules corresponding to the $\theta = 15^{\circ}$ and 0$^\circ$ situations are exactly the same. In both scenarios, only the \textDelta$_{\bar{4},4}$ ($k=0$) and \textDelta$_{0,4}$ ($k=4$) gaps are non-zero, while the other $k$-even QTM gap, \textDelta$_{\bar{2},4}$ ($k=2$), vanishes when $H_{\text{T}}=0$. These results imply that the easy-axis tilting does not affect the symmetry of the Hamiltonian, contrary to the case for the Mn$_3$ SMM. This can be understood in terms of the different symmetry properties associated with $q$-even and $q$-odd cases. In the even case, the molecular \textit{z}-axis must also be a $C_2$ axis. Consequently, forcing the local $C_2$ axes of the individual ions to be parallel to the molecular $z$-axis ($\theta = 0^{\circ}$) does not introduce an extra $C_2$ symmetry to the molecular Hamiltonian. In contrast, the molecular \textit{z}-axis is not a $C_2$ axis in a molecule with odd rotational symmetry. Therefore, the symmetry of the molecular Hamiltonian changes when $\theta = 0^\circ$.

In the preceding discussions of Mn$_3$, the QTM selection rules can be simply understood in terms of the rotational symmetry of the molecule ($C_6$ or $C_3$). In contrast, the {selection rules} for Ni$_4$ cannot be fully explained by the $S_4$ molecular symmetry; one must additionally take into account the intrinsic $C_i$ symmetry of the spin Hamiltonian. Upon application of a magnetic field parallel to the molecular \textit{z}-axis, the $S_4$ symmetry group reduces to $C_2$, for which the \textDelta$_{\bar{2},4}$ ($k=2$) QTM resonance should be allowed. This clearly contradicts the simulation in Fig.\,\ref{fig:Ni4Gap}, which suggests a higher symmetry. However, one must also consider the $C_i$ symmetry associated with the SO interaction. The consequential zero-field spin Hamiltonian then possesses $S_4 \times C_i = C_{4h}$ symmetry, which corresponds to the symmetry of the $\hat{O}_4^4$ interaction, as seen in Fig.\,\ref{fig:GSAHighOrder}d. Upon application of a longitudinal field, the $C_{4h}$ group reduces to the $C_4$ group, for which the expected QTM selection rule $|\text{\textDelta}m|=4n$ is recovered. The $C_i$ symmetry is guaranteed by the nature of the {SO interaction}. This property is not limited to spin Hamiltonians, \textit{i.e.}, it applies to any Hamiltonian dictated by {crystal field} and/or SO physics, where the $C_i$ symmetry should apply regardless of whether the orbital angular momentum is quenched or not. In other words, it is always necessary to consider the $C_i$ symmetry in addition to the structural symmetry, especially when {improper rotations} are involved. Unfortunately, observation of $k > 0$ QTM steps in Ni$_4$ is impractical due to the extremely fast tunneling at $k = 0$. This tunneling should be greatly suppressed if the ground spin state of the molecule is increased, as is of course the case for Mn$_{12}$. However, it would be interesting to obtain a four-fold symmetric SMM constituted of four $s = 2$ Mn\textsuperscript{III} ions, for which it would be possible to study the $k > 0$ QTM steps. Moreover, the Hamiltonian dimension of just $625\times625$ would be quite manageable.

\subsection{{Disorder}}
\label{sec:DisorderNi4}

\begin{figure}[b]
\centering
\includegraphics*[width=.7\textwidth]{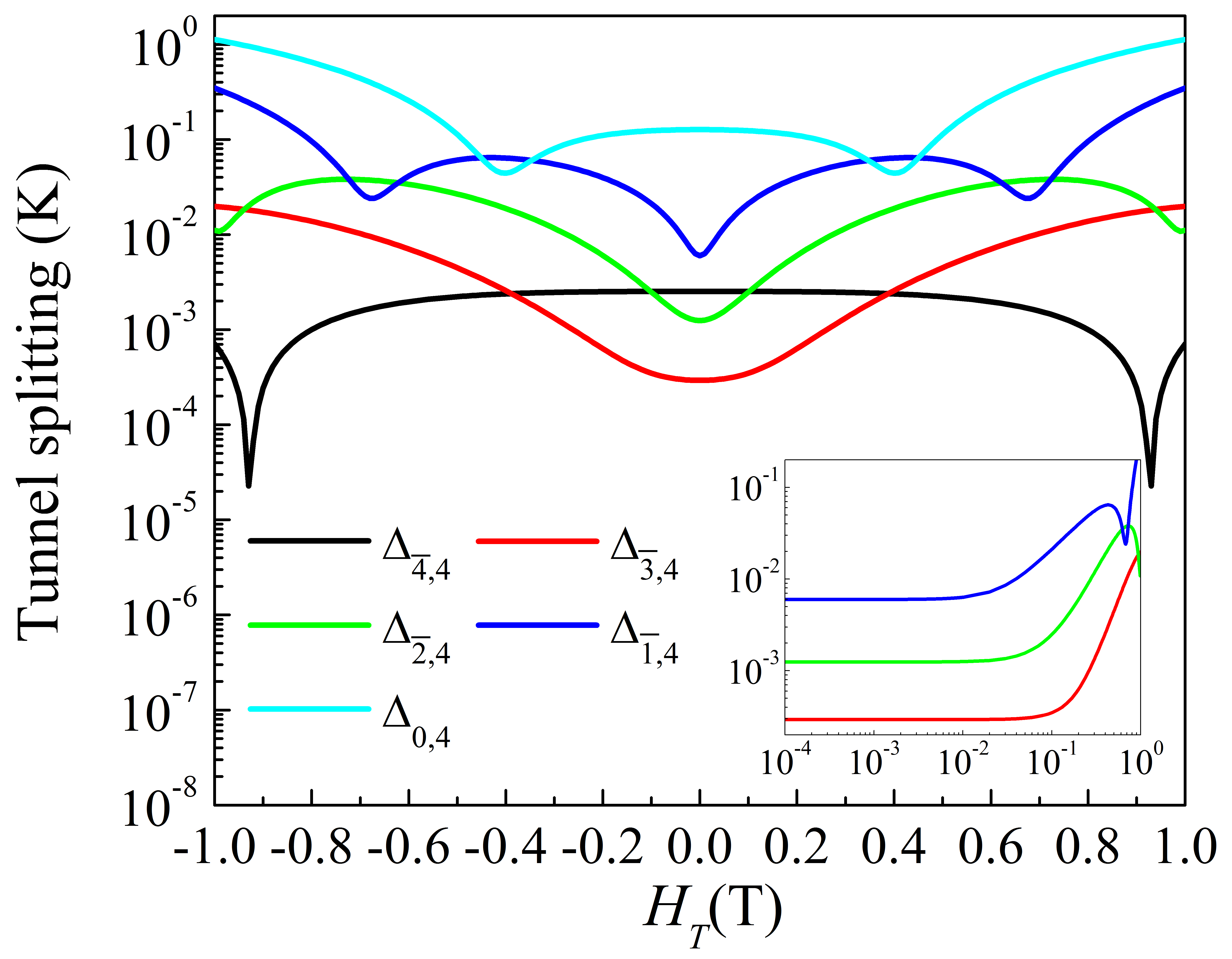}
\caption[]{The effect of disorder on the ground state QTM gaps for the Ni$_4$ SMM. The simulations were performed employing Eqn.\,\ref{eqn:Ni4MS} by misaligning the zfs tensor of one of the Ni$^{\text{II}}$ ions with respect to the unperturbed molecular $z$-axis (see details in the main text).}
\label{fig:Ni4Disorder}
\end{figure}

In the presence of random disorder, one would expect the symmetry of most molecules to be lowered, leading to an absence of QTM selection rules. The Ni$_4$ molecule provides an excellent platform to study this issue. Fig.\,\ref{fig:Ni4Disorder} was generated by adjusting the orientation of the zfs tensor of one of the Ni\textsuperscript{II} ions in the molecule, \textit{i.e.}, the zfs tensors of three of the Ni ions are tilted $15^{\circ}$ from the molecular \textit{z}-axis, while the other is tilted $10^{\circ}$. It should be emphasized that it is not trivial to find the orientation of the molecular easy-axis in this situation, \textit{i.e.}, it no longer coincides with the molecular $z$-axis. For each resonance ($k=0$ to 4), a search was performed for the minimum QTM gap by varying the orientation of the applied field. As seen in the figure, all resonances posses a non-zero QTM gap at $H_{\text{T}}=0$. The inset to Fig.\,\ref{fig:Ni4Disorder} plots \textDelta$_{\bar{3},4}$ ($k=1$), \textDelta$_{\bar{2},4}$ ($k=2$) and \textDelta$_{\bar{1},4}$ ($k=3$) on a log-log scale, clearly demonstrating that these QTM gaps, which are forbidden for $S_4$ symmetry, also saturate at non-zero values when $H_{\text{T}}\rightarrow 0$. These results show that a small disorder effectively unfreezes all QTM steps without the assistance of a transverse field. This argument can be reinforced by {group theoretic considerations}. With random disorder, the symmetry of a molecule is lowered to $C_1$, resulting in a spin Hamiltonian with $C_i$ symmetry. Upon applying a longitudinal field, the $C_i$ group reduces to $C_1$, where all of the states necessarily belong to the same one-dimensional irreducible representation \cite{Koster1963MITpress}. Therefore, mixing between all states is allowed. We note that this kind of disorder can be introduced by small crystallographic defects, which always exist to some degree in real samples. Thus, exceptionally clean crystals are required in order to observe symmetry imposed QTM selection rules. Importantly, the preceding discussion clearly demonstrates that {disorder} can be responsible for the observation of $k$-odd QTM steps in SMMs with even rotational symmetries.

\section{{Quantum Tunneling of Magnetization} in Low-Symmetry {Mn$_4$} Single-Molecule Magnets}
\label{sec:Mn4}

In order to contrast results presented in previous sections, {EPR} and QTM/{BPI} results are presented here for two related Mn$_4$ SMMs that possess almost no symmetry. Both molecules crystallize in the triclinic P$\bar{1}$ space group. One of the structures co-crystallizes with solvent, while the other does not. Consequently, significant differences are observed in terms of the widths of EPR and QTM resonances due to the different degrees of disorder in the two crystals. In addition, small structural differences associated with the Mn$_4$ cores result in different coupling strengths between the Mn ions within the two molecules which, in turn, result in different QTM behavior. 

\subsection{The Mn$_4$ Single-Molecule Magnets}
\label{sec:Mn4SMMs}

The Mn$_4$ molecules (Figs.\,\ref{fig:Mn4molecules}a and b) possess mixed-valent butterfly-type structures, with two central Mn\textsuperscript{III} ions ($s_2=s_4=2$) in the body positions and two Mn\textsuperscript{II} ions  ($s_1=s_3=5/2$) in the wing positions (see Fig.\,\ref{fig:Mn4molecules}c). Magnetic superexchange is mediated through oxygen bridges. The Mn$_4$ molecule that co-crystallizes with solvent is {[Mn$_4$(anca)$_4$(Hedea)$_2$(edea)$_2$]$\cdot$2CHCl$_3\cdot$2EtOH}, henceforth Mn$_4$-anca (see Fig.\,\ref{fig:Mn4molecules}a and Ref. \cite{Liu2011} for more details). The solvent-free molecule is {[Mn$_4$(Bet)$_4$(mdea)$_2$(mdeaH)$_2$](BPh$_4$)$_4$}, henceforth Mn$_4$-Bet (see Fig.\,\ref{fig:Mn4molecules}b and Ref.\cite{Heroux2011} for more details). Both molecules crystallize in the centrosymmetric triclinic space group P$\bar{1}$. The asymmetric unit therefore consists of half the molecule (Mn\textsuperscript{III}Mn\textsuperscript{II}), with the other half generated via an inversion. This also ensures that all four Mn ions lie in a plane.  Temperature dependent susceptibility data recorded at different magnetic fields indicate that both molecules possess a spin $S=9$ ground state at low temperatures, implying overall ferromagnetic coupling within the molecules (note that this does not rule out the possibility that one of the exchange paths could be antiferromagnetic).

\subsection{{EPR} and QTM Spectroscopy in {Mn$_4$} SMMs with and without Solvent}
\label{sec:QTMMn4}

A selection of EPR and QTM measurements from Refs.\cite{Liu2011,Heroux2011} are presented in Fig.\,\ref{fig:Mn4data}: (a) displays 165 GHz EPR spectra recorded at different temperatures for a single crystal of Mn$_4$-anca, with the magnetic field applied close to the molecular easy-axis; (b) shows equivalent spectra obtained for Mn$_4$-Bet at a frequency of 139.5 GHz and similar temperatures. The first thing to note are the obvious differences in the EPR linewidths in the two figures. This again provides a clear illustration of the inferior quality of samples in which the SMMs co-crystallize with interstitial solvent molecules. In the present example, the Mn$_4$-anca sample is the more disordered, resulting in a broader distribution of GSA zfs parameters. A series of nine absorption peaks can clearly be seen for the Mn$_4$-anca SMM in Fig. \,\ref{fig:Mn4data}a, which have been labeled 1 through 9. These correspond to transitions from consecutive spin projection ($m_S$) states belonging to the $S=9$ ground state multiplet, where the numbering denotes the absolute $m_S$ value associated with the level from which the EPR transition was excited, \textit{e.g.}, resonance $\alpha=9$ corresponds to the $m_S=-9\rightarrow -8$ transition. The fact that all of the spectral weight transfers to the $\alpha=9$ resonance as $T\rightarrow 0$ indicates uniaxial anisotropy, \textit{i.e.}, $D<0$ according to the GSA Hamiltonian of Eqn.\,\ref{eqn:GSA2nd}. The uneven spacing between the labeled ground state resonances is indicative of {4\textsuperscript{th}- (and higher-) order anisotropy} within the GSA (or {weak exchange} within the MS picture). Finally, as the temperature is increased, a few weaker resonances (not labeled) can be seen to appear in Fig. \,\ref{fig:Mn4data}a between the labeled transitions. These additional peaks are associated with the population of higher-lying, {excited spin states}, \textit{e.g.}, $S=8$.

The EPR spectra obtained for Mn$_4$-Bet (Fig. \,\ref{fig:Mn4data}b) are not so simple to interpret. First and foremost, many more peaks are observed, suggesting the population of many more spin states. Based upon the knowledge gained from Mn$_4$-anca, and the results of subsequent simulations, the nine resonances corresponding to transitions within the $S=9$ ground state multiplet have been labeled in the figure. Meanwhile, the peaks that are not labeled correspond to transitions within low-lying excited spin multiplets. Clearly, the emergence of excited state EPR transitions at much lower temperatures indicates significantly weaker exchange coupling in the Mn$_4$-Bet molecule; note that excited state intensity is seen even at the lowest temperature, whereas this is not the case until $\sim9$~K in the Mn$_4$-anca sample. The weaker exchange and higher crystal quality associated with the Mn$_4$-Bet sample lead to the observation of unusual MQT/BPI behavior, as detailed in the following section.

\begin{figure}[t]
\centering
\includegraphics*[width=.7\textwidth]{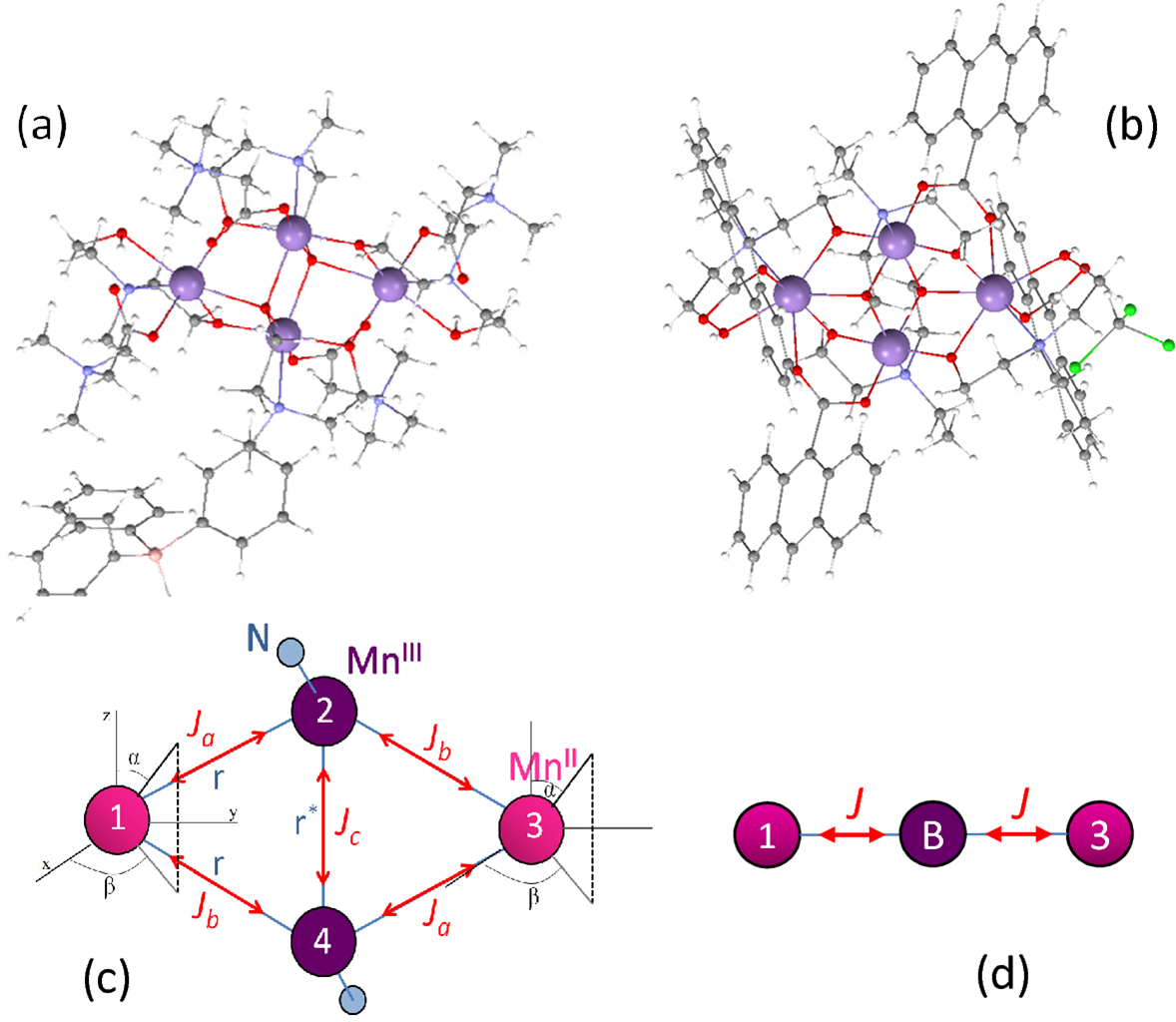}
\caption[]{(a) The Mn$_4$-anca molecule. (b) The Mn$_4$-Bet molecule. Color code: Mn = purple, O = red, N = blue, C = grey, H = white, B = pink and Cl = green. (c) Sketch showing the different exchange interactions used to solve the four spin MS Hamiltonian for these molecules. (d) Trimer model representing the Mn$_4$ molecules assuming an infinite $J$ coupling between the two central Mn$^{\rm III}$ ions.}
\label{fig:Mn4molecules}
\end{figure}

\begin{figure}[b]
\centering
\includegraphics*[width=1\textwidth]{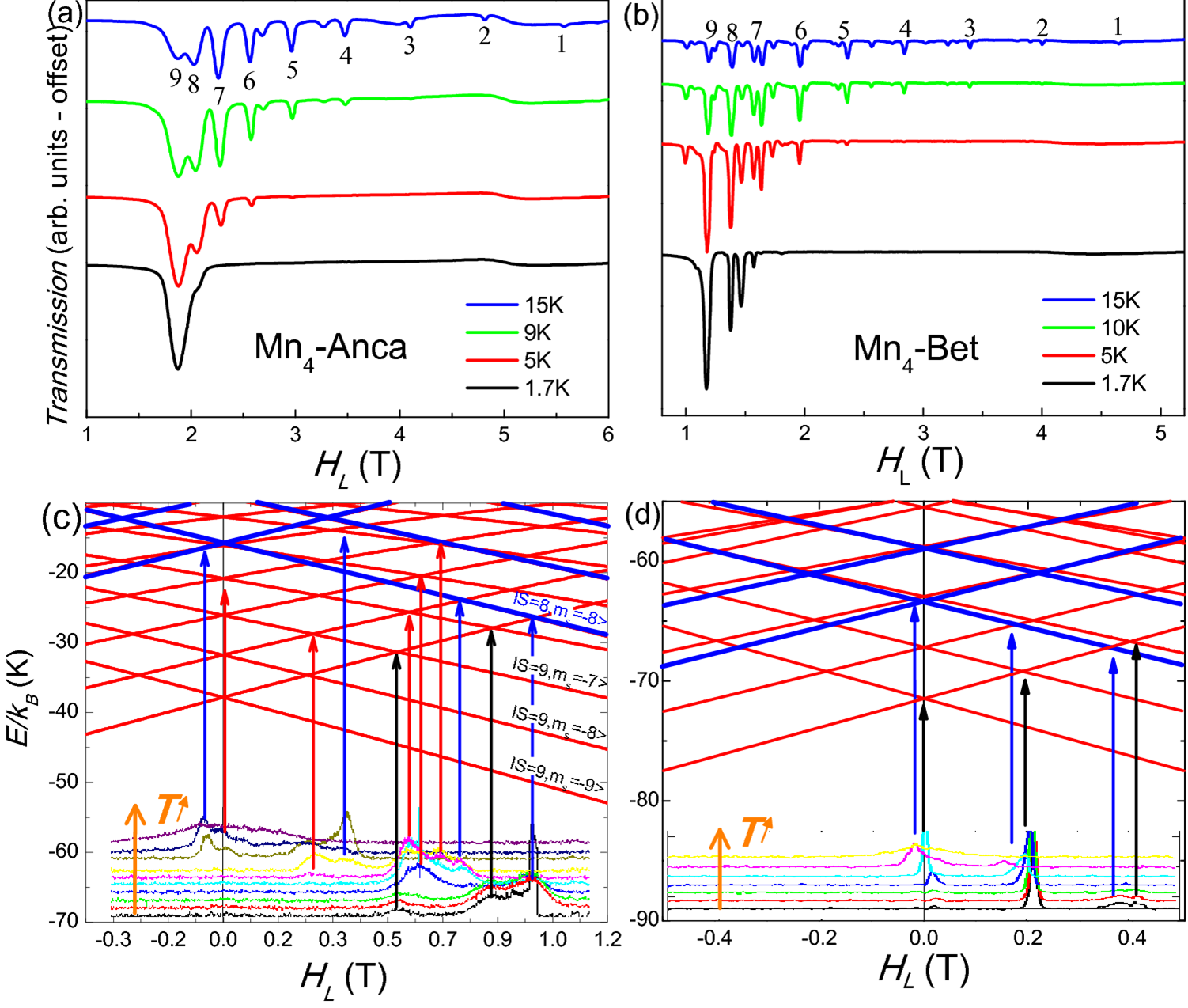}
\caption[]{EPR spectra obtained at different temperatures with the field along the easy-axis at: 165 GHz for Mn$_4$-anca (a); and 139.5 GHz for Mn$_4$-Bet (b). Zeeman diagrams depicting the low-lying energy levels for Mn$_4$-anca (c) and Mn$_4$-Bet (d), obtained by diagonalization of the MS Hamiltonian (Eqn\,\ref{eqn:MS}) using the trimer model depicted in Fig.\,\ref{fig:Mn4molecules}d. Arrows relate the QTM peaks observed in the field derivatives of the magnetization curves obtained at different temperatures (bottom of the graphics) with the corresponding crossings between spin levels; with black arrows indicating crossings of the ground state $|S=9,m_s=9\rangle$ (red for excited states) with other states within the same spin multiplet ($S=9$), and blue arrows signaling both ground and excited crossings involving levels of different spin length ($S=9$ and $S=8$).}
\label{fig:Mn4data}
\end{figure}

Although one can reproduce the peak positions of the nine labeled EPR transitions in both Figs. \,\ref{fig:Mn4data}a and b using the GSA (including $p\geq4$ terms), a MS Hamiltonian becomes essential to account for transitions within excited spin states. In other words, one starts to see the limitations of the GSA in these two cases$\--$especially for Mn$_4$-Bet. Diagonalization of the exact MS Hamiltonian that considers all four Mn ions and the couplings between them (as indicated in Fig.\,\ref{fig:Mn4molecules}c) is manageable on a standard computer. However, a convenient and reasonable approximation replaces the ferromagnetically coupled central Mn\textsuperscript{III} ions with a single $s_B=4$ spin, resulting in a linear {trimer} consisting of the central spin, $s_B$, and the two outer $s_A=5/2$ Mn\textsuperscript{II} spins, as depicted in Fig.\,\ref{fig:Mn4molecules}d. This approximation, which contains elements of both the GSA and MS models, has been compared with the more exact four-spin model for the case of Mn$_4$-Bet in Ref.~\cite{Quddusi2011}. The two models give good agreement in terms of the lowest-lying portions of the energy-level diagrams that dictate the low-temperature QTM and EPR properties, provided that the ferromagnetic coupling between the Mn\textsuperscript{III} ions is not too weak. Although it involves a level of approximation, the trimer model captures all of the important physics associated with these low-symmetry SMMs. Moreover, the smaller Hamiltonian matrix dimension enables much faster fitting (hours instead of days), and employs fewer parameters$\--$just two \textit{d} tensors and a single exchange coupling constant, \textit{J}. Finally, working with a single $J$ parameter to identify the internal exchange energy becomes particularly useful in relating the distinct behavior of the two molecules to different intramolecular couplings which result from slight structural disparities between the two compounds. In the following, we diagonalize the MS Hamiltonian (Eqn.\,\ref{eqn:MS}) using the trimer model (Fig.\,\ref{fig:Mn4molecules}c) to account for the energy landscape associated to the lowest lying molecular spin multiplets, which result from the main anisotropy terms in the Hamiltonian (i.e. axial terms). The full MS Hamiltonian (Eqn.\,\ref{eqn:MS}) including the four manganese ions (Fig.\,\ref{fig:Mn4molecules}d) is used to account for the behavior of the tunnel splittings, which result from the smaller anisotropy terms in the Hamiltonian (i.e. transverse terms) and are more sensitive to small variations of the internal degrees of freedom of the molecules.

In fitting the data in Figs.\,\ref{fig:Mn4data}a and \,\ref{fig:Mn4data}b, as well as other EPR data obtained at different temperatures and applied field orientations (see Refs.\cite{Liu2011,Heroux2011}), one finds that the {exchange coupling} constant $J$ has a strong influence on the positions of the EPR peaks (particularly the relative spacings between peaks). This again highlights the fact that one cannot use a GSA to realistically describe these results, especially those of Mn$_4$-Bet, \textit{i.e.}, there is no exchange parameter in the GSA (all energy splittings are parameterized in terms of the SO anisotropy). It is thus preferable to use the MS approach whenever computational resources will allow, as is the case for all of the low-nuclearity SMMs described in this chapter. Indeed, there are many other interesting features associated with the magnetic behavior of SMMs that cannot be explained with the GSA, regardless of how complex the corresponding Hamiltonian is, \textit{e.g}., QTM resonances involving level crossings between different spin states.

QTM spectroscopy also facilitates comparisons between the two Mn$_4$ molecules. Figs.\,\ref{fig:Mn4data}c and d show the association between the observed QTM resonance positions and the corresponding {energy-level diagrams} for the Mn$_4$-anca and Mn$_4$-Bet molecules, respectively. The {QTM resonances} are determined from the positions of the peaks in the derivatives of the magnetization versus field curves obtained at different temperatures, as shown in the lower portions of Figs.\,\ref{fig:Mn4data}c and d. Note that the effects of the {solvent disorder} can again be seen, causing broader QTM resonances for Mn$_4$-anca in comparison to Mn$_4$-Bet. The energy level diagrams are obtained via exact diagonalization of the MS Hamiltonian of Eqn.\,\ref{eqn:MS} using the trimer model depicted in Fig.\,\ref{fig:Mn4molecules}c with the following parameters: $J=5.42$ K, $d_1=d_2=d_A=-0.115$ K, $d_B=-2.22$ K in Mn$_4$-anca (Fig.\,\ref{fig:Mn4data}c); and $J=1.90$ K, $d_A=-0.115$ K, $d_B=2.00$ K in Mn$_4$-Bet (Fig.\,\ref{fig:Mn4data}d), with isotropic $g=2.0$ for all ions in both cases. The correspondence between QTM resonances and their associated level crossings are indicated by vertical arrows. Crossings involving the ground state $|S=9,m_s=9\rangle$ with other levels within the same multiplet ($S=9$) are indicated by black arrows and expected to appear at the lowest temperatures, for which transitions involving all excited states (red arrows) should vanish. Interestingly, some of the resonances (blue arrows) correspond to crossings between levels associated with different spin states, \textit{i.e}., $S=9$ and $S=8$. The main difference between the two molecules resides in the value of $J$, being more than double for Mn$_4$-anca. This results in the lowest spin projection states ($m_S=\pm8$) associated with the first excited state ($S=8$) being much closer to the $m_S=\pm 9$ ground states in Mn$_4$-Bet ($\sim8$~K separation) than in Mn$_4$-anca ($\sim22$~K separation). These findings are consistent with the temperature dependence of the EPR spectra, which suggested the excited states to be appreciably lower in energy in the Mn$_4$-Bet molecule in comparison to Mn$_4$-anca. The differences in $J$ values can be reconciled with the minor structural differences between the two molecules. It is well known that the superexchange coupling between two transition metal ions is very sensitive to the bridging angle, to the extent that the sign of the interaction can switch from ferromagnetic to antiferromagnetic within a small range of angles \cite{Brechin2009Dalton1,Brechin2009Dalton2,Hill2010Dalton}. Indeed, there are measurable differences in the bond angles and distances associated with these two Mn$_4$ molecules.

\subsection{{Berry Phase Interference} in {Mn$_4$}-Bet}
\label{sec:BPIMn4}

The spectroscopic results presented in the previous section illustrate how small structural perturbations can lead to significant changes in the exchange coupling within a molecule. Crucially, the Mn$_4$-Bet SMM resides in a particularly interesting region of the `anisotropy' versus `exchange' parameter space in which {excited spin states} exert a significant influence on the low-energy/low-temperature quantum properties. First and foremost, it can be seen that some of the QTM resonances involve level crossings between different spin multiplets. More importantly, the QTM properties within this intermediate exchange regime ($d\sim J$) are extraordinarily sensitive to the internal magnetic structure of the molecule. As noted in previous sections, the physics associated with the strong exchange limit ($J>>d$) is dominated by the 2\textsuperscript{nd}-order GSA anisotropy. Consequently, any observable BPI patterns should display a high degree of symmetry ($D_{2h}$), regardless of the molecular symmetry. However, in the intermediate exchange regime, one may expect any BPI effects to mimic the symmetry of the molecule under investigation much more closely. This symmetry can be expressed exactly using a four spin MS model in the case of Mn$_4$-Bet, although we note that one can also reproduce some of the features of these experiments using a GSA that includes appropriate {4\textsuperscript{th}- (and higher-) order terms}. The virtual lack of any symmetry associated with Mn$_4$-Bet leads to some remarkable BPI patterns, which also shed light on a previous mystery surrounding another high-nuclearity Mn$_{12}$ SMM~\cite{delBarco2010PRB}.

\begin{figure}[t]
\centering
\includegraphics*[width=1\textwidth]{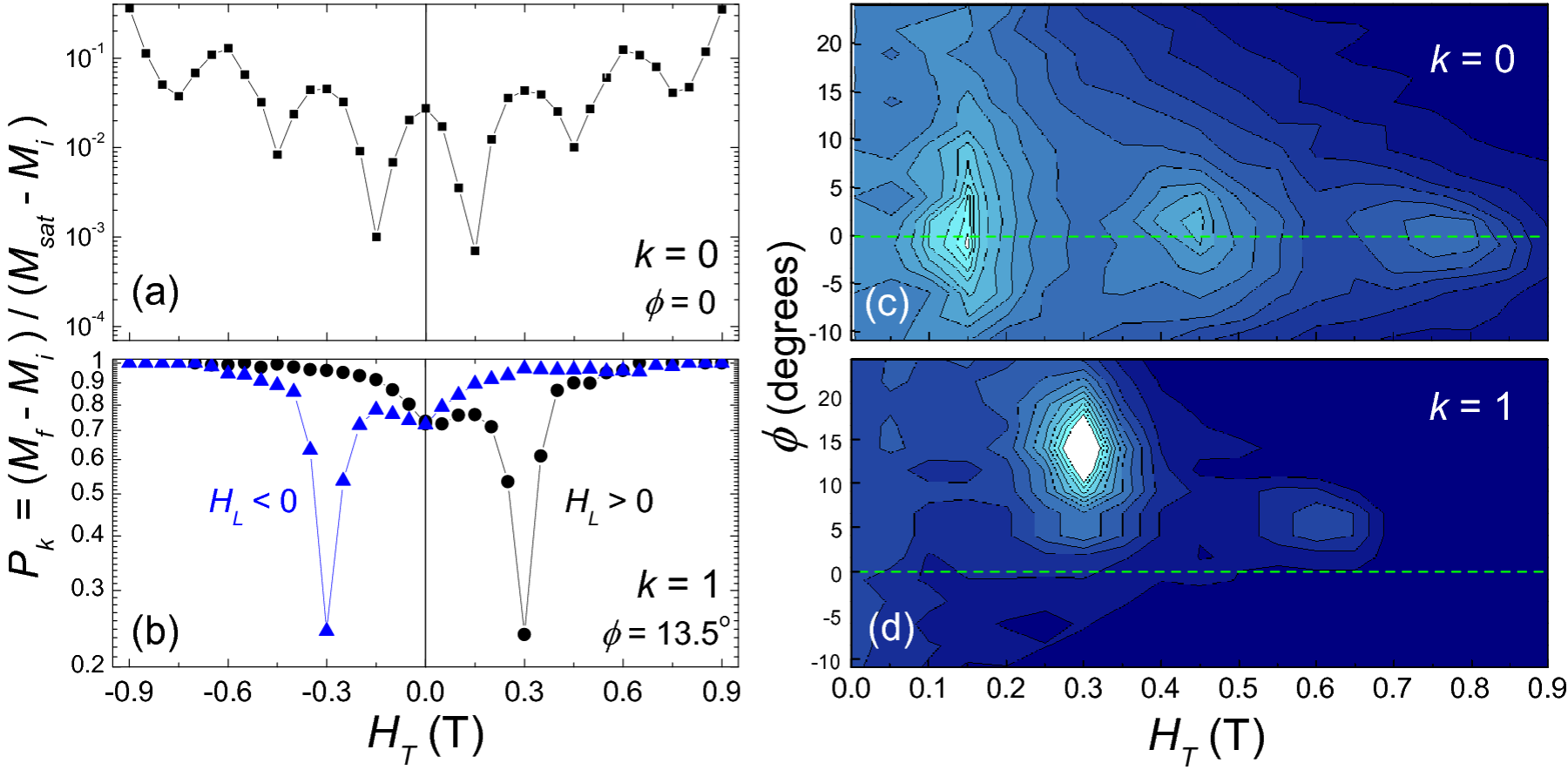}
\caption[]{Modulation of the QTM probabilities for resonances $k=0$ (a) and $k=1$ (b) as a function of $H_T$ applied at different angles, $\phi$, within the $xy$-plane of the Mn$_4$-Bet SMM. The asymmetry of the BPI pattern of oscillations in resonance $k=1$ is inverted upon reversal of $H_L$. (c) and (d) show the modulation of the tunnel splitings of resonances $k=0$ and $k=1$, respectively, for different directions of the transverse field.}
\label{fig:Mn4BPI}
\end{figure}

Low temperature QTM measurements performed in the presence of a transverse magnetic field, $H_{\text{T}}$, enable exploration of the dominant symmetries associated with the Hamiltonian describing a SMM. Fig.\,\ref{fig:Mn4BPI}a shows the modulation of the QTM probability for the $k=0$ resonance for Mn$_4$-Bet as a function of the magnitude of $H_{\text{T}}$ applied parallel to the hard axis of the molecule ($\phi=0^\circ$, see Ref.\cite{Quddusi2011} for details). As explained in Sec.\,\ref{sec:BPIMn3}, the oscillations correspond to {BPI} (constructive/destructive interference associated with equivalent tunneling trajectories on the Bloch sphere), with minima occurring at regularly spaced field values ($\text{\textDelta}H_{\text{T}}=0.3$ T). Experiments designed to probe the modulation of the $k=0$ QTM gap as a function of the orientation of a fixed transverse field within the hard plane (see Ref.\cite{Quddusi2011}) reveal a two-fold behavior. One may be tempted to ascribe this to a rhombic anisotropy. However, the molecule possesses a much lower symmetry (P$\bar{1}$). In fact, the two-fold pattern is a direct manifestation of the $C_i$ symmetry associated with the SO interaction. Since no longitudinal field ($H_L$) is present for the $k=0$ resonance, the Hamiltonian must be invariant with respect to inversion of $H_{\text{T}}-$hence the apparent two-fold behavior. Note that the $k=0$ BPI oscillations do, indeed, respect the symmetry of the Hamiltonian, \textit{i.e.}, they are invariant under inversion of $H_{\text{T}}$.

Due to the absence of $H_{\text{L}}$, $k=0$ turns out \textit{not} to be the most interesting QTM resonance, because the $C_i$ symmetry guarantees symmetric BPI patterns about $H_{\text{T}}=0$. In contrast, this is clearly not the case for the BPI pattern associated with the $k=1$ resonance, as can be seen in Fig.\,\ref{fig:Mn4BPI}b. In this case, a single interference minimum is observed at $H_{\text{T}}=0.3$~T for only one polarity of the transverse field, \textit{i.e.}, the corresponding BPI minimum is completely absent under inversion of $H_{\text{T}}$. Such a result is not so surprising when one recognizes that there are no mirror symmetries within {P$\bar{1}$}. Hence, there is no reason why the BPI patterns should be invariant under inversion of just one component of the applied field. However, the Hamiltonian, and therefore the BPI patterns, must be invariant under a full inversion of the applied field, \textit{i.e.}, inversion of both $H_{\text{T}}$ and $H_{\text{L}}$ together. This indeed turns out to be the case for the $k=1$ resonance, as clearly seen in Fig.\,\ref{fig:Mn4BPI}b.

Another interesting feature observed in the BPI patterns of Mn$_4$-Bet is the motion of the minima associated with the $k=1$ resonance within the $xy$-plane. This can be observed in Fig.\,\ref{fig:Mn4BPI}d, which shows a color contour plot of the QTM probability in $k=1$, as a function of the magnitude and the orientation of $H_{\text{T}}$. Two minima can clearly be observed; they are again spaced by $\sim0.3$~T, and are located half way between the $k=0$ minima, as is usually the case for even/odd resonances. However, the $k=1$ minima do not appear at the same orientations within the $xy$-plane as those of $k=0$. Moreover, the orientations of the two observed $k=1$ minima do not even coincide: $\phi=13.5^\circ$ for the first minimum and $\phi=6^\circ$ for the second one. Note that, in contrast to $k=1$, all of the $k=0$ minima lie along the nominal hard anisotropy axis ($\phi=0$), as seen in Fig.\,\ref{fig:Mn4BPI}c. In essence, the hard directions associated with the $k=1$ resonance (for which both $H_{\text{L}}$ and $H_{\text{T}}$ are finite) do not occur along a fixed axis, as would be the case for a rhombic molecule. This property, which is analogous to the behavior seen in Fig.\,\ref{fig:Mn3PRL}b, is a direct consequence of the absence of any mirror symmetries in the P$\bar{1}$ space group. It is impossible to simultaneously satisfy both $C_i$ symmetry and a mirror symmetry if the BPI minima do not reside on a fixed axis in space. However, if the mirror symmetry is broken, then the BPI minima may in principle occur anywhere on the Bloch sphere, so long as they occur in pairs related by inversion. The results displayed in Fig.\,\ref{fig:Mn4BPI}d were obtained by rotating $H_{\text{T}}$ while keeping $H_{\text{L}}$ fixed; the space above and below the $xy$-plane was not explored due to experimental constraints. One cannot rule out further minima above or below the $xy$-plane. Indeed, this may explain why the 2\textsuperscript{nd} $k=1$ minimum is so weak, \textit{i.e.}, its real location may be above or below the $xy$-plane. In fact, this could also be true for the $k=0$ minima, as again emphasized in Fig.\,\ref{fig:Mn3PRL}b for the case of the Mn$_3$ molecule, which lacks \textit{xy} mirror symmetry. More detailed (and time consuming) experiments are clearly required to further explore this issue in Mn$_4$-Bet.

In order to simulate the observed BPI patterns, one must obviously break some or all of the {mirror symmetries} within the Mn$_4$-Bet spin Hamiltonian, whilst also respecting the {inversion symmetry} of the real molecule. There really is only one way to achieve this, involving misalignment of the 2\textsuperscript{nd}-order zfs tensors associated with the Mn$^{\rm III}$ and the Mn$^{\rm II}$ ions. The molecular inversion symmetry guarantees that the JT axes associated with the Mn$^{\rm III}$ ions be parallel to each other; likewise the zfs tensors associated with the Mn$^{\rm II}$ ions must be co-linear. However, there is no requirement that the tensors associated with the two types of ion be co-linear. Indeed, all of the results in Fig.\,\ref{fig:Mn4BPI} have been reproduced in Ref.\cite{Quddusi2011} following exactly this approach. Although the trimer model (Fig.\,\ref{fig:Mn4molecules}c) can reproduce the observed behavior quite well, the four spin Hamiltonian (Eqn.\,\ref{eqn:MS}) was employed in order to describe the geometry in Fig.\,\ref{fig:Mn4molecules}d, since it gives a better quantitative agreement and allows for a more physical interpretation of the observations (e.g. the real dipolar coupling between the four Mn ions can be employed, which involves no fitting parameters). Using this approach, we find the optimal parameter set to be as follows: (central Mn$^{\rm III}$'s) $d_2=d_4=-4.99$~K and $e_2=e_4=0.82$~K, with the easy and hard axes along $z$ ($\alpha_2=0$) and $x$ ($\beta_2=0$), respectively; (outer Mn$^{\rm II}$'s) $d_1=d_3=-0.67$~K and $e_1=e_3=0$, with the axes rotated with respect to the central spin by identical Euler angles $\alpha_{1,3}=45^\circ$, $\beta_{1,3}=0^\circ$ (as required by inversion symmetry); $\gamma$ being zero for all ions; finally, isotropic ferromagnetic exchange constants $J_a=-3.84$~K, $J_b=-1.20$~K and $J_c=-3.36$~K were employed. It should be stressed that these parameters were additionally constrained by the EPR and QTM data in Fig.\,\ref{fig:Mn4data}. They do not necessarily constitute the correct parameterization, but they account for all experimental observations. Note that the zfs tensors of the two Mn$^{\rm II}$ ions are tilted by 45 degrees with respect to the Mn$^{\rm III}$ tensors, thereby breaking the molecular \textit{xy} mirror symmetry. This results in a breaking of the \textit{xy} mirror symmetry of the corresponding spin Hamiltonian. Note that this would not be the case for a molecule with even rotational symmetry, because of the $C_i$ symmetry associated with the SO interaction. However, in P$\bar{1}$ ($q=1$), the \textit{xy} mirror symmetry of the Hamiltonian is broken, as was also the case for the trigonal ($q=3$) Mn$_3$ SMM. One could, in principle, also explain these results using a GSA by introducing 4$^{\rm th}$- (and higher-) order terms. However, a more natural and satisfying account of the results is obtained by diagonalizing the four-spin Hamiltonian, which easily allows for a tilting of the zfs tensors of the four Mn ions. 

Interestingly, there again exists a connection to Mn$_{12}$, albeit a wheel molecule that bears no resemblance to the well studied, high-symmetry {Mn$_{12}$}'s discussed in Sec.\,\ref{sec:disorderMn3}. The {Mn$_{12}$ wheel} molecule~\cite{delBarco2010PRB}, which possesses the exact same {P$\bar{1}$ symmetry} as the Mn$_4$ molecules considered in this section, attracted considerable controversy on account of the observation of asymmetric $k>0$ {BPI patterns}~\cite{Ramsey2008NaturePhys,Wernsdorfer2008PRL}. Initial attempts to account for this behavior involved treating the molecule as a {dimer}, including an unphysical {Dzyaloshinskii-Moriya} coupling between the two halves of the dimer (this interaction is forbidden on account of the molecular inversion symmetry~\cite{delBarco2009PRL}). The present studies have shown that asymmetric $k>0$ BPI patterns are, in fact, quite natural in low-symmetry molecules. As in the case of Mn$_3$, a detailed understanding of the QTM characteristics in the Mn$_4$ molecules is made possible due to the low nuclearity of the system, which enables the employment of a MS Hamiltonian. This, in turn, provides fundamental insights that are much harder to achieve when studying larger molecules using a GSA.

\section{Summary and Outlook}
\label{sec:summary}

This chapter focuses on the microscopic factors that dictate the QTM behavior observed in polynuclear transition-metal SMMs, with particular focus on molecular symmetry. The examples provided involve relatively simple, low-nuclearity clusters (Mn$_{3}$, Mn$_{4}$ and Ni$_{4}$) which display essentially the same physics as the original Mn$_{12}$ and Fe$_8$ SMMs that have occupied chemists and physicists working in this field for nearly two decades. The simpler systems are amenable to analysis using a microscopic spin Hamiltonian that incorporates both the single-ion physics, and isotropic exchange coupling between the constituent ions, and relies on relatively few parameters. One can therefore systematically investigate the role of internal spin degrees of freedom within a molecule, in contrast to the approximate GSA approach employed for most studies of Mn$_{12}$ and Fe$_8$. Comparisons between theory and experiment are presented for a range of cluster symmetries, with remarkable quantitative agreement achieved in all cases.

Until fairly recently, most SMM research was directed towards polynuclear $3d$ transition metal clusters, with the synthetic goal of maximizing both the molecular spin state and the magneto-anisotropy. However, a number of fundamental factors have limited progress based on this strategy, with the record blocking temperature for a Mn$_6$ cluster \cite{Milios2007JACS} only just surpassing that of the original Mn$_{12}$ SMM \cite{Redler2009PRB}. Limiting factors include: (i) a tendency for superexchange interactions between constituent transition metal spins to be both weak (few cm$^{-1}$) and often antiferromagnetic; (ii) the fact that strong crystal-field effects typically quench the orbital momentum associated with $3d$ elements, thus significantly suppressing the magneto-anisotropy; and (iii) the difficulties associated with maximally projecting any remaining (2\textsuperscript{nd} order SO) anisotropy onto the ground spin state. In fact, careful studies of this issue suggest that one is unlikely to achieve anisotropy barriers that significantly exceed those of the constituent ions \cite{Hill2010Dalton}. This is perhaps best illustrated by the optimum Mn$_3$, Mn$_6$ and Mn$_{12}$ SMMs, which possess similar barriers (to within a factor of $< 2$ \cite{Hill2010Dalton}). This is because the molecular anisotropy, $D$, is given by a weighted sum of the anisotropies of the constituent ions ($d_i$), where the weighting is inversely proportional to the total molecular spin, $S$ \cite{Hill2010Dalton,WaldmannIC2007}. Thus, $D$ decreases as $S$ increases, and the theoretical best that one can hope to achieve is an anisotropy barrier ($\sim DS^2$) that scales linearly with $S$ (or $N$, the number of magnetic ions in a ferromagnetic molecule). Even then, many challenges remain$\--$some fundamental (quantum tunneling, spin state mixing, \textit{etc}. \cite{Hill2009PRB}), some synthetic. The synthetic challenges, in particular, become more complex with increasing molecule size, \textit{e.g.}, maintaining ferromagnetic couplings, maintaining parallel arrangements of the individual anisotropy tensors, \textit{etc}. Thus, it is perhaps no surprise that the optimum [Mn\textsuperscript{III}]$_N$ SMM has a nuclearity of just six \cite{Milios2007JACS}!

Given the aforementioned situation, it has become clear that a more direct route to magnetic molecules that might one day be used in practical devices involves the use of ions that exhibit considerably stronger magnetic anisotropies than those that have traditionally been employed in the synthesis of large polynuclear clusters, \textit{i.e}., ions for which the orbital momentum is not quenched, and/or heavier elements in which strong SO effects are manifest. Examples include certain high-symmetry and low-coordinate $3d$ transition metal complexes (Fe\textsuperscript{II} \cite{HarmanJACS2010}, Co\textsuperscript{II} \cite{ZadroznyChemComm2012}, even Ni\textsuperscript{II} \cite{RuampsJACS2012}), as well as elements further down the periodic table such as the $4f$ and $5f$ elements. Indeed, over the past few years, a number of mononuclear complexes have been shown to exhibit magnetization blocking of pure molecular origin \cite{IshikawaJACS2003,AlDamenJACS2008,RinehartJACS2009,HarmanJACS2010,ZadroznyChemComm2012}. However, the quantum magnetization dynamics of these single-ion molecular nanomagnets has yet to be studied in detail, and much remains to be learned theoretically. Obviously, much of the physics related to exchange which is discussed in this chapter does not apply in these cases. Nevertheless, the spin Hamiltonian formalism remains applicable, as does the crucial importance of molecular and crystallographic symmetry. In particular, 4\textsuperscript{th} and higher-order crystal-field interactions may be expected to play a crucial role in the quantum dynamics of mononuclear lanthanide SMMs \cite{GhoshDalton2012}. Thus, one would expect similar combinations of theory and spectroscopy to contribute in future to this evolving field of research.

\section{Acknowledgements}
\label{sec:acknowl}

This work was supported by the US National Science Foundation, grant numbers DMR0804408 (SH), CHE0924374 (SH), and DMR0747587 (EdB). Work performed at the National High Magnetic Field Laboratory is supported by the National Science Foundation (grant number DMR1157490), the State of Florida and the Department of Energy.

\clearpage

\begin{table}[t]
\centering
\caption[]{Comparison of tunneling gaps obtained for Mn$_3$ from the MS and GSA models for resonances $k$ = 0, 1, 2 and 3, for the two cases $\theta=0$ (top) and $\theta=8.5^{\circ}$ (bottom).}
\renewcommand{\arraystretch}{1.2}
\setlength\tabcolsep{5pt}
\begin{tabular}{@{}ccccccc@{}}
\hline\noalign{\smallskip}
$k$ & $n$ & \textDelta & GSA-gap (K) & MS-gap (K) & Ratio \\
\hline\noalign{\smallskip}
\multicolumn{6}{c}{JT-axes parallel to the molecular z-axis} \\
\hline\noalign{\smallskip}
0 & 2 & \textDelta$_{\bar{3},3}$ & $2.60\times 10^{-2}$ & $2.66\times 10^{-2}$ & 0.98 \\
0 & 2 & \textDelta$_{\bar{6},6}$ & $1.10\times 10^{-6}$ & $1.05\times 10^{-6}$ & 1.05 \\
2 & 2 & \textDelta$_{\bar{2},4}$ & $2.37\times 10^{-2}$ & $2.35\times 10^{-2}$ & 1.01 \\
\hline\noalign{\smallskip}
\multicolumn{6}{c}{JT-axes tilted $\theta = 8.5^{\circ}$ away from the molecular z-axis} \\
\hline\noalign{\smallskip}
0 & 2 & \textDelta$_{\bar{3},3}$ & $2.76\times 10^{-2}$ & $2.91\times 10^{-2}$ & 0.95 \\
0 & 4 & \textDelta$_{\bar{6},6}$ & $1.26\times 10^{-6}$ & $1.25\times 10^{-6}$ & 1.01 \\
1 & 3 & \textDelta$_{\bar{4},5}$ & $4.68\times 10^{-5}$ & $4.19\times 10^{-5}$ & 1.12 \\
1 & 1 & \textDelta$_{\bar{1},2}$ & $6.33\times 10^{-2}$ & $6.31\times 10^{-2}$ & 1.00 \\
2 & 2 & \textDelta$_{\bar{2},4}$ & $2.45\times 10^{-2}$ & $2.61\times 10^{-2}$ & 0.94 \\
3 & 3 & \textDelta$_{\bar{3},6}$ & $8.66\times 10^{-5}$ & $7.53\times 10^{-5}$ & 1.15 \\
3 & 1 & \textDelta$_{0,3}$ & $1.76\times 10^{-1}$ & $1.76\times 10^{-1}$ & 1.00 \\
\hline
\end{tabular}
\label{Table:Mn3mapping}
\end{table}

\clearpage

\clearpage

\clearpage

\clearpage

\clearpage

\clearpage

\clearpage

\clearpage

\clearpage

\printindex
\end{document}